\documentclass[12pt,a4paper]{article}

\usepackage[utf8]{inputenc}
\usepackage[T1]{fontenc}
\usepackage{amsmath}
\usepackage{amssymb}
\usepackage{xspace}
\usepackage{ae}
\usepackage{color}
\usepackage[affil-it]{authblk}
\usepackage{graphicx}
\usepackage{bbm}
\usepackage{bm}
\usepackage{enumerate}
\usepackage{cite}

\newcommand{\rd}{\ensuremath{\mathrm{d}}}

\newcommand{\bw}{\bm{w}}
\newcommand{\bu}{\bm{u}}
\newcommand{\bx}{\bm{x}}

\newcommand{\R}{\mathcal{R}}
\newcommand{\ads}{AdS$_3$\xspace}
\newcommand{\slr}{SL(2,$\mathbf{R}$)\xspace}
\newcommand{\LS}{$\mathcal{L}$\xspace}

\usepackage[colorlinks=true, linkcolor=blue, bookmarks=true]{hyperref}

\begin{document}

\title{Black hole formation from pointlike particles in three-dimensional anti-de Sitter space}

\author[1,2]{Jonathan Lindgren\thanks{Electronic adress: \texttt{Jonathan.Lindgren@vub.ac.be}}}
\affil[1]{Theoretische Natuurkunde, Vrije Universiteit Brussel, and the International Solvay Institutes, Pleinlaan 2, B-1050 Brussels, Belgium}
\affil[2]{Physique Th\'eorique et Math\'ematique, Universit\'e Libre de Bruxelles, Campus Plaine C.P.\ 231, B-1050 Bruxelles, Belgium}
\date{}

\maketitle

\abstract{We study collisions of many point-like particles in three-dimensional anti-de Sitter space, generalizing the known result with two particles. We show how to construct exact solutions corresponding to the formation of either a black hole or a conical singularity from the collision of an arbitrary number of massless particles falling in radially from the boundary. We find that when going away from the case of equal energies and discrete rotational symmetry, this is not a trivial generalization of the two-particle case, but requires that the excised wedges corresponding to the particles must be chosen in a very precise way for a consistent solution. We also explicitly take the limit when the number of particles goes to infinity and obtain thin shell solutions that in general break rotational invariance, corresponding to an instantaneous and inhomogeneous perturbation at the boundary. We also compute the stress-energy tensor of the shell using the junction formalism for null shells and obtain agreement 
with the point particle picture.}
\newpage

\tableofcontents
\section{Introduction}



Black hole formation in anti-de Sitter space has recently attracted a lot of interest due to the AdS/CFT correspondence. The reason is that according to the AdS/CFT dictionary, black hole formation in anti-de Sitter space is supposed to be dual to the equilibration into a thermal state in the dual field theory. Such processes are difficult to study using conventional field theory techniques and AdS/CFT thus provides a remarkable tool to explore strongly coupled field theories out of equilibrium, a field known as ``holographic thermalization''. Black hole formation in anti-de Sitter space and asymptotically anti-de Sitter spaces has been investigated in many different setups, and the equilibration of many different field theory observables has been studied, see for instance \cite{Balasubramanian:2011ur,Keranen:2015fqa,Balasubramanian:2013oga,Chesler:2008hg,Bhattacharyya:2009uu,Bizon:2011gg,Liu:2013iza}.\\

Even though there are no black hole solutions in three dimensions with zero cosmological constant, we do have a black hole solution when the cosmological constant is negative, the so called BTZ black hole\cite{Banados:1992wn}. This solution has very similar properties to higher dimensional black holes, and has proven to be a very useful toy model to study quantum and thermodynamical aspects of black holes. There is also an interesting way to create solutions corresponding to dynamical formation of BTZ black holes, which was first discussed in \cite{Matschull:1998rv}. In this setup two massless pointlike particles are created at the boundary of three-dimensional anti de Sitter space (\ads), falling radially into the bulk, and then colliding to form a joint object. Pointlike particles are here defined as conical singularities, and described by excising a wedge of \ads. Depending on the energy of the two particles, the result is either a massive pointlike particle or a BTZ black hole. Three dimensions is 
special since pointlike particles do not collapse into black holes themselves, as opposed to in higher dimensions, and a negative cosmological constant is required if we want to be able to interpret the resulting solution as a black hole. This construction also relies on the fact that three-dimensional gravity has no local degrees of freedom and all solutions are thus locally equivalent to \ads (but not globally), so we do not expect this to work in higher dimensions. These solutions were originally studied due to the general interest in having access to analytical toy models for black hole formation, but they are of course also interesting in light of the \ads/CFT$_2$ correspondence.\\

Other constructions using pointlike particles in \ads have also been explored. In \cite{Holst:1999tc} the case of two particles with non-zero impact parameter was discussed (two particles missing each other), and the resulting spacetime was interpreted as a rotating BTZ black hole. This spacetime could also be interpreted as an anti-de Sitter analog of the Gott universe\cite{Gott:1990zr}, a spacetime containing closed time-like curves, thus establishing a connection between the Gott universe and the rotating BTZ black hole. In \cite{DeDeo:2002yg}, an eternal time machine was discovered, with two pointlike particles orbiting each other forever, eternal in the sense that there are no event or chronology horizons and the closed time-like curves fill the entire space. Recently this time-machine was studied in the context of the AdS/CFT correspondence \cite{Arefeva:2015sza}. The boundary dual to a single moving conical defect as well as the dual of two colliding particles have also been studied\cite{Arefeva:
2015zra,Ageev:2015qbz,Ageev:2015xoz,Balasubramanian:1999zv}.\\

In this paper we study the formation of a BTZ black hole or of a massive pointlike particle from the collision of an arbitrary number of massless particles created at the boundary, thus generalizing the result in \cite{Matschull:1998rv}. We show how to construct this solution for particles falling on arbitrary angles and with arbitrary energies, which is not a trivial generalization of the two-particle case but requires a more clever construction where the wedges excised behind the particles take a very precise form (they are in general not the standard symmetric wedges used in \cite{Matschull:1998rv}). We then proceed to take the limit of infinitely many particles, and show that this will give rise to a thin shell spacetime that in general breaks rotational symmetry. We also analyze these solutions using the thin shell formalism for null shells\cite{Barrabes:1991ng}, and obtain perfect agreement. Namely, the stress-energy tensor computed from this formalism is that of a pressureless static fluid with an 
energy that depends on the angular coordinate, and is exactly proportional to the distribution of the massless particles. These solutions are also very interesting in their own right, since they would correspond in the dual field theory to an injection of energy that has an angular dependence, and would allow us to study inhomogeneous holographic thermalization processes, similar to what was studied in \cite{Balasubramanian:2013oga} in four spacetime dimensions. However, we do not expect to find such simple inhomogeneous thin shell solutions in higher dimensions, since the dynamical modes of the metric are expected to be excited.\\

The paper is organized as follows: In Section \ref{secAds} we introduce three-dimensional anti-de Sitter space and the relevant coordinate systems. In Section \ref{secParticle} we study in detail the (boosted) massive point-like particle as well as the massless particle corresponding to the limit when the boost parameter goes to infinity. In Section \ref{secBTZ} we review the BTZ black hole solution and construct it in a general boosted coordinate system, which will be useful when interpreting the result of the collision as a black hole. In Section \ref{collisions} we discuss colliding particles, starting by reviewing the two-particle case, and then proceeding with the construction of $n$ colliding particles, explaining in particular why this is a non-trivial generalization of the two-particle case. In Section \ref{geometry} we show how to do a coordinate transformation to explicitly see that the geometry after the collision corresponds to a massive pointlike particle or to a BTZ black hole. In Section \ref{secLimit} we take the limit $n\rightarrow\infty$ and obtain thin shell solutions which in general break rotational symmetry. In Section \ref{thinshell} we compute the stress-energy tensor of these thin shell solutions using the junction conditions for null shells, following the algorithm in \cite{Barrabes:1991ng} and \cite{Musgrave:1997}. In Section \ref{secConclusions} we end with some conclusions and future research directions. Some technical derivations have been put in the appendices.


\section{Three dimensional anti-de Sitter space}\label{secAds}
Anti-de Sitter space in three dimensions can be defined via an embedding in four dimensional Minkowski space with signature $(-1,-1,1,1)$. The embedding equation is
\begin{equation}
x_3^2+x_0^2-x_1^2-x_2^2=1,\label{embedding}
\end{equation}
and the ambient space has the metric
\begin{equation}
\rd s^2=\rd x_1^2+\rd x_2^2-\rd x_3^2-\rd x_0^2.
\end{equation}
This can be covered by for example the following coordinate system
\begin{equation}
\begin{array}{cc}
x^3=\cosh \chi \cos t, & x^0=\cosh \chi \sin t,\\
x^1=\sinh \chi \cos \phi, & x^2=\sinh \chi \sin \phi, \label{adscoord}
\end{array}
\end{equation}
where $t$ is a time coordinate, $\phi$ and angular coordinate and $\chi$ a radial coordinate. Note that in this embedding the time coordinate $t$ would be periodic and we would have closed time-like curves. $AdS_3$ is thus defined by the {\it covering space} of this manifold, effectively dropping the periodicity of $t$ (or ``unwinding'' the manifold). The coordinate ranges are thus $0\leq\chi\leq\infty$, $0\leq\phi<2\pi$ and $-\infty< t<\infty$. The metric in these coordinates is given by
\begin{equation}
\rd s^2=\rd \chi^2+\sinh^2\chi\rd \phi^2-\cosh^2\chi\rd t^2.\label{adschimetric}
\end{equation}
Another useful coordinate system can be obtained by the radial reparametrization
\begin{equation}
r=\tanh(\chi/2),
\end{equation}
where $0\leq r\leq 1$, for which the metric becomes
\begin{equation}
\rd s^2=\left(\frac{2}{1-r^2}\right)^2(\rd r^2+r^2\rd \phi^2)-\left(\frac{1+r^2}{1-r^2}\right)^2\rd t^2.\label{adsmetric}
\end{equation}
This is a particularly useful coordinate system, since at fixed time $t$ the geometry is that of a Poincar\'e disc, and space-like geodesics at constant $t$ can be visualized as circle segments\cite{Matschull:1998rv}. We will use coordinate system \eqref{adsmetric} for all figures (thus mapping every static slice to a disc), but \eqref{adschimetric} when doing computations.\\

Another useful property that we will use in this article is that $AdS_3$ is equivalent to the covering group of \slr, which is the group of real matrices with unit determinant. The basis of matrices that we will use is the unit matrix along with 
\begin{equation}
\gamma_0=\left(\begin{array}{ccc} 0 & 1 \\ -1 & 0 \\ \end{array}\right),\mathrm{ }\gamma_1=\left(\begin{array}{ccc} 0 & 1 \\ 1 & 0 \\ \end{array}\right),\mathrm{ } \gamma_2=\left(\begin{array}{ccc} 1 & 0 \\ 0 & -1 \\ \end{array}\right).
\end{equation}
If we expand a matrix $\bx=x_3\mathbf{1}+\gamma_a x^a$, where raising of indices are done with the matrix $\eta_{ab}=diag(-1,1,1)$, then the condition of unit determinant becomes exactly the embedding equation \eqref{embedding}. The isometries of \ads can now be obtained by left and right multiplications of elements in \slr
\begin{equation}
\bx\rightarrow \bm{g}\bx\bm{h}^{-1} \hspace{20pt} \bm{g},\bm{h}\in \textrm{\slr}.
\end{equation}
Geodesics can now be obtained as one parameter subgroups in \slr, generated by a single group element, and the metric coincides with the killing metric in \slr. For more details we refer the reader to \cite{Matschull:1998rv}. \\

It will also be convenient to define the following matrices
\begin{equation}
\omega(\alpha)=\cos \alpha + \sin\alpha\gamma_0, \hspace{20pt}\gamma(\alpha)=\cos \alpha\gamma_1 + \sin\alpha\gamma_2,
\end{equation}
so that an element in \slr can be written as
\begin{equation}
\bx=\cosh\chi\omega(t)+\sinh\chi\gamma(\phi).
\end{equation}
We also have the following useful relations
\begin{equation}
\begin{array}{cc}
\gamma(\alpha)\gamma(\beta)=\omega(\alpha-\beta),& \gamma(\alpha)\omega(\beta)=\gamma(\alpha-\beta),\\
\omega(\alpha)\omega(\beta)=\omega(\alpha+\beta),&\omega(\alpha)\gamma(\beta)=\gamma(\alpha+\beta).\\
\end{array}
\end{equation}

\section{Pointlike particles as conical singularities}\label{secParticle}

In three dimensions, pointlike particles are equivalent to conical singularities moving on either time-like or light-like geodesics. All properties of the particle are stored in their {\it holonomy}, which is an element in SL(2,R) corresponding to an isometry of \ads. This is the isometry for which we identify space when going around the particles world line. In particular, a stationary massive particle at the origin of \ads, with angle deficit $2\nu$ (meaning that the total angle around the singularity is $2\pi-2\nu$), can be described by cutting out a wedge enclosed by two planes $w_\pm$ of constant angles $\phi_\pm$ such that $\phi_+-\phi_-=2\nu$. These planes are thus mapped to each other by a rotation of angle $2\nu$ which is described as the group element $\omega(\nu)=e^{\gamma_0\nu}=\cos \nu + \gamma_0\sin \nu\in$SL(2,R), which is then the holonomy of the particle. Thus we have $\omega(\nu)w_-=w_+\omega(\nu)$. The mass is given by the mass shell relation as $\frac{1}{2}\mathrm{Tr} \bu = \cos m$ which 
gives $m=\nu$\cite{Matschull:1997du}. This is a particular coordinate system for describing a spacetime obtained by identifying points in \ads by $\bx\rightarrow\omega(\nu)^{-1}\bx\omega(\nu)$.\\

It is also possible to consider the patch between $\phi_-$ and $\phi_+$ as the allowed spacetime, instead of the removed spacetime. This would correspond to a pointlike particle with mass $m=\pi-\nu$. A more general coordinate system can be obtained considering $n$ wedges such that $w_+^i$ is mapped to $w_-^{i+1}$, and the mass would then be $m=\pi-\sum_i\nu_i$ (or equivalently, removing $n$ wedges such that the mass is $m=\sum_i\nu_i$). This is a rather trivial construction, but we will use it later on.
\subsection{Boosted particle}
A moving massive pointlike particle can now be obtained by boosting the stationary one. Let us first only look at how the world line of the particle transforms under the boost. The stationary particle is given by the world line
\begin{equation}
\mathbf{x}=\cos \tau+\sin \tau \gamma_0,
\end{equation}
which corresponds to the origin $\chi=0$. We can boost this by acting with the isometry $\bx'=\bu^{-1}\bx\bu$ where
\begin{equation}
\bu=e^{-\frac{1}{2}\zeta\gamma(\alpha)}=\cosh \frac{1}{2}\zeta-\gamma(\alpha)\sinh \frac{1}{2}\zeta.
\end{equation}
We have
\begin{equation}
\begin{split}
\bu^{-1}\bx\bu&=\left(\cosh \frac{1}{2}\zeta+\gamma(\alpha)\sinh \frac{1}{2}\zeta\right)\left(\cos t+\sin t \gamma_0\right)\left(\cosh \frac{1}{2}\zeta-\gamma(\alpha)\sinh \frac{1}{2}\zeta\right)\\
&=\cos t+\cosh^2\frac{1}{2}\zeta \sin t \gamma_0+2\sinh\frac{1}{2}\zeta\cosh\frac{1}{2}\zeta \sin t\gamma(\alpha-\pi/2)+\sin t \sinh^2\frac{1}{2}\zeta\gamma_0\\
&=\cos t+\cosh\zeta \sin t \gamma_0-\sinh \zeta \sin t\gamma(\alpha+\pi/2)
\end{split}
\end{equation}
This is now a particle oscillating around the origin at an angle of $\psi\equiv\alpha+\pi/2$. In terms of the new coordinates $(\chi',t',\phi')$ defined by $\bx'=\bu^{-1}\bx\bu$ one can show that the geodesic satisfies
\begin{equation}
\tanh\chi'=-\tanh \zeta \sin t'
\end{equation}
Note that for negative $\sin t'$ (assuming positive $\zeta$), $\chi'$ is positive. For positive $\sin t'$ one can either just abuse the coordinate system and allow for negative $\chi'$ or insist on positive $\chi'$ and change the angle by $\pi$.\\

Let us now consider the equation for the lines $w_\pm$, which bound the removed wedge, after a general boost. Applying the group element $\bu$ gives

\begin{equation}
\begin{split}
\bu^{-1}x\bu&=\left(\cosh \frac{1}{2}\zeta+\gamma(\alpha)\sinh \frac{1}{2}\zeta\right)\\
&\left(\cosh \chi \cos t + \cosh \chi \sin t \gamma_0 + \sinh \chi \gamma(\phi)\right)\left(\cosh \frac{1}{2}\zeta-\gamma(\alpha)\sinh \frac{1}{2}\zeta\right)\\
&=\cos t\cosh\chi+\cosh\chi\cosh\zeta \sin t \gamma_0+\cosh\chi\sinh \zeta \sin t\gamma(\alpha-\pi/2)\\
&+\cosh^2\frac{1}{2}\zeta\sinh \chi \gamma(\phi)-\sinh\chi\sinh^2 \frac{1}{2}\zeta\gamma(2\alpha-\phi)+\sinh \chi\sinh \zeta\sin(\alpha-\phi)\gamma_0
\end{split}
\end{equation}

\begin{figure}
\includegraphics[scale=0.7]{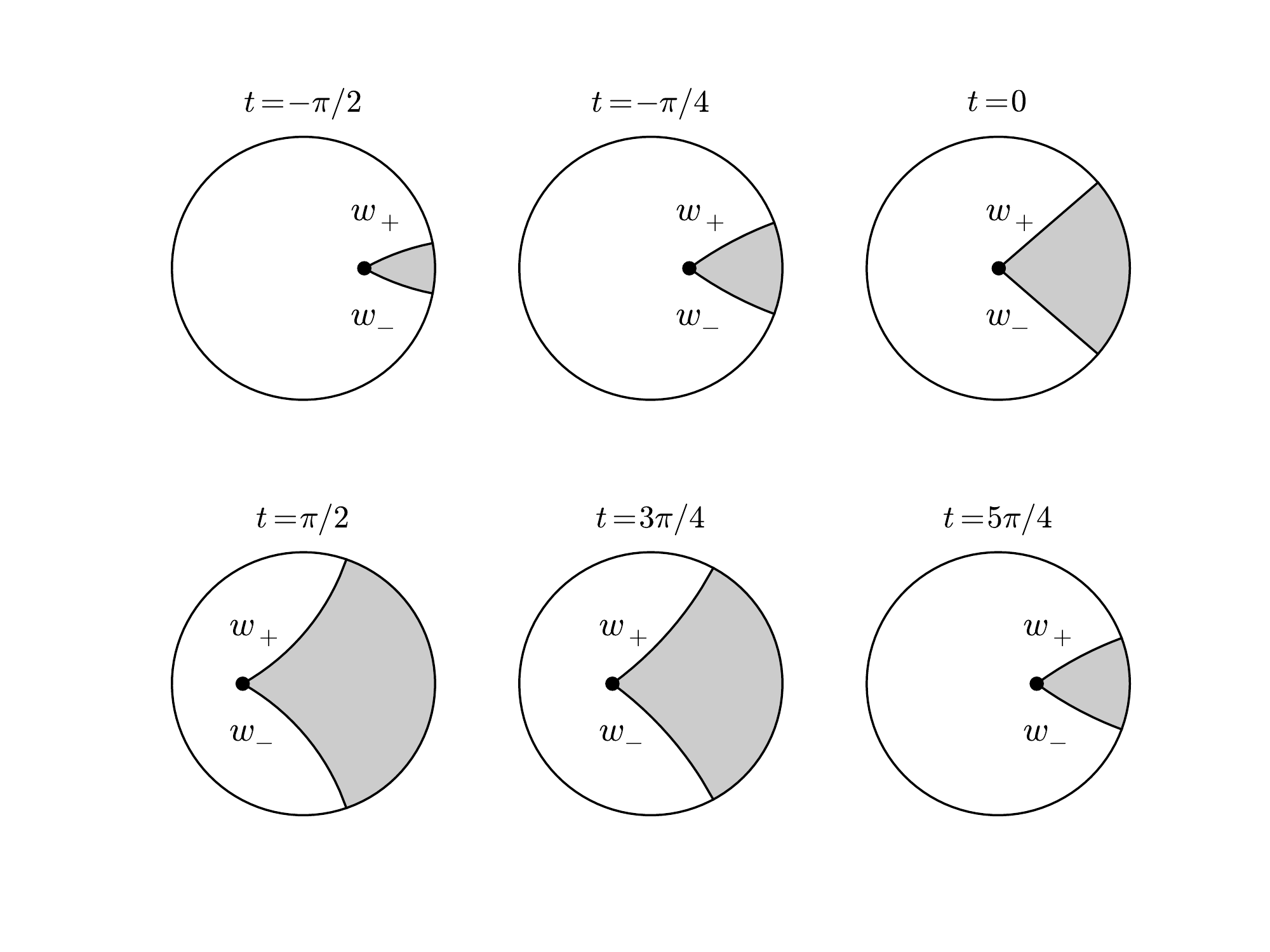}
\caption{\label{1massive}A boosted massive particle in \ads. The particle follows a boosted timelike geodesic, and a symmetric wedge had been excised behind the particle to account for the interaction with gravity according to equation \eqref{c1}, with $p=0$ (symmetric wedge), $\zeta=1$ and $\nu=1/2$. The dark grey regions are removed parts of spacetime, the white regions are the allowed space. The surfaces $w_+$ and $w_-$ are identified, meaning that when we pass from the white region past $w_\pm$, we are mapped to $w_\mp$. }
\end{figure}

In the new coordinates this is described by $(\chi',\phi',t')$ which gives the system of equations
\begin{align}
\begin{split}
\cos t'\cosh \chi'&=\cos t\cosh \chi,\\
\sin t'\cosh \chi'&=\cosh\chi\sin t \cosh \zeta + \sinh\chi\sinh\zeta\sin(\alpha-\phi),\\
\sinh\chi'\cos\phi'&=\cosh\chi\sinh\zeta\sin t\sin\alpha+\sinh\chi(\cosh^2\frac{1}{2}\zeta\cos\phi-\sinh^2\frac{1}{2}\zeta\cos(2\alpha-\phi)),\\
\sinh\chi'\sin\phi'&=-\cosh\chi\sin t\sinh\zeta\cos\alpha+\sinh\chi(\cosh^2\frac{1}{2}\zeta\sin\phi-\sinh^2\frac{1}{2}\zeta\sin(2\alpha-\phi)).\label{boosteqs}
\end{split}
\end{align}
The planes bordering the removed wedge for a stationary massive particle are given by constant angles $\phi_\pm$, so we would like to find an equation relating the new coordinates $\chi'$, $t'$ and $\phi'$ with $\phi_\pm$, $\zeta$ and $\psi=\alpha+\pi/2$ as parameters. For simplicity we redefine $\phi_{\pm}=\psi\pm(1\pm p)\nu$ such that the total deficit angle is $2\nu$, and $p$ then parametrizes how much $\phi_\pm$ deviates from being located symmetrically around $\psi$. It turns out (see Appendix \ref{app1}) that the $w_\pm^i$ in the new coordinate system are determined by the equations
\begin{equation}
\tanh\chi'\sin(-\phi'+\Gamma_\pm+\psi)=-\tanh\zeta\sin\Gamma_\pm \sin t'. \label{c1}
\end{equation}
where
\begin{equation}
\tan\Gamma_\pm=\pm\tan((1\pm p)\nu)\cosh\zeta.\label{Gammapm}
\end{equation}
There is an ambiguity of $\pi$ in $\Gamma_\pm$, and it will be chosen such that it goes to zero as $\nu\rightarrow0$. However, the formula \eqref{c1} should be interpreted with care. For $\sin t'<0$ we have for $w_\pm$ that
\begin{equation}
\phi'\in(\psi_i,\arcsin(\sin t \sin \Gamma_\pm)+\psi_i+\Gamma_\pm).
\end{equation}
However, for $\sin t'>0$, the range is instead
\begin{equation}
\phi'\in(-\arcsin(-\sin t \sin \Gamma_\pm)+\psi_i+\Gamma_\pm,\psi_i\pm\pi),
\end{equation}
while for $\sin t'=0$ we have $\phi'=\psi+\Gamma_\pm$. A boosted massive pointlike particle is shown in Figure \ref{1massive}.\\
Let us now compute the holonomy for the boosted particle. The holonomy of the stationary particle was given by $\omega(\nu)$. In the new coordinates, given by $\bx'=\bu^{-1}\bx \bu$ where $\bu=e^{-\frac{1}{2}\zeta\gamma(\alpha)}$, the new surfaces bordering the removed patch $w'_\pm$ will be related by
\begin{equation}
w'_-=\bw'^{-1}w'_+\bw'
\end{equation}
where
\begin{equation}
\bw'=\bu^{-1}\omega(\nu) \bu
\end{equation}
which then represents the holonomy of the moving particle. Evaluating this we have
\begin{align}
\bw'=&\cos \nu+\gamma_0\cosh \zeta \sin \nu-\sinh \zeta \sin \nu \gamma(\psi).\label{hol}
\end{align}

\subsection{The massless limit}
\begin{figure}
\includegraphics[scale=0.7]{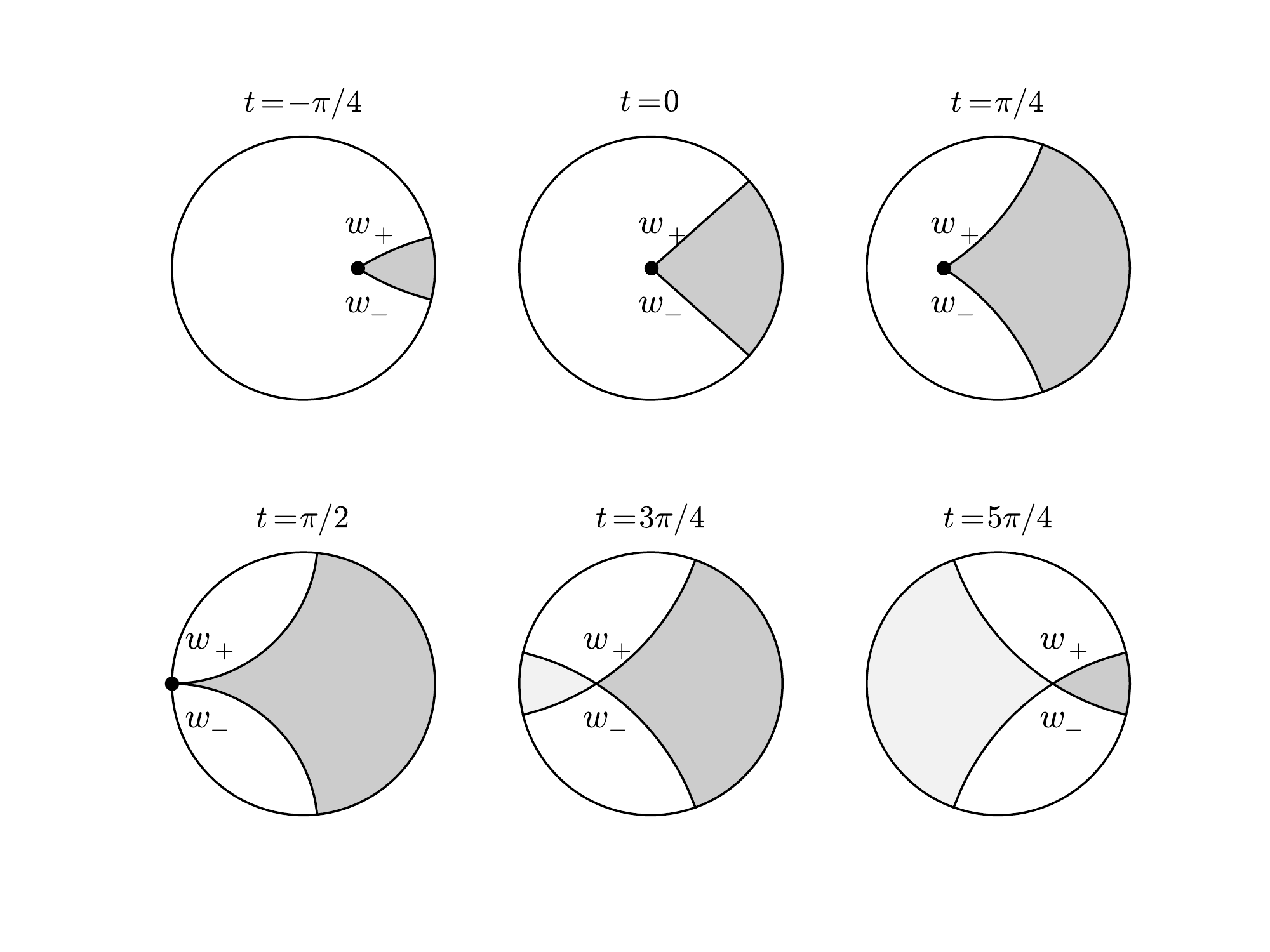}
\caption{\label{1massless}A massless particle moving through \ads, with a symmetric wedge excised behind the particle. The dark grey regions are removed parts of spacetime, the white regions are the allowed space and the light shaded region is just two patches of allowed geometry overlapping each other. The surfaces $w_+$ and $w_-$ are identified, meaning that when we pass from the white region past $w_\pm$, we are mapped to $w_\mp$. The particle is annihilated at $t=\pi/2$, and after the annihilation the spacetime is that of empty \ads.}
\end{figure}

We will use the result for the boosted massive particle to construct a massless particle. This corresponds to taking the boost parameter to infinite while letting the mass go to zero, namely by taking the limit $\zeta\rightarrow\infty$, $\nu\rightarrow0$ such that $\cosh\zeta\sin\nu\rightarrow\tan\epsilon\equiv e$. Using the expression \eqref{hol} we obtain that
\begin{equation}
\bu'=1+\tan\epsilon(\gamma_0-\gamma(\psi)),
\end{equation}
which is the holonomy of a massless particle moving through \ads along a radial geodesic at angle $\psi$, consistent with the expressions in \cite{Matschull:1998rv}. We will mostly use the parameter $e$ instead of the parameter $\epsilon$. The full spacetime corresponding to the moving massless particle is obtained by taking this limit in equations \eqref{c1} and \eqref{Gammapm}, to obtain
\begin{equation}
\tanh\chi'\sin(-\phi'+\Gamma_\pm+\psi)=-\sin\Gamma_\pm\sin t',\label{masslessp}
\end{equation}
where
\begin{equation}
\tan\Gamma_\pm=\pm(1\pm p)\tan\epsilon=\pm(1\pm p)e.\label{Gammapmmassless}
\end{equation}
The special case of $p=0$ corresponds to wedges located symmetrically behind the particle, which was used in \cite{Matschull:1998rv}. It should be emphasized here that the physical parameters for this massless pointlike particle are $\psi$ (the angular position of the radial geodesic) and $e=\tan\epsilon$ (which is related to the energy of the particle), while $p$ just parametrizes our coordinate system and has no physical meaning (but this freedom will be crucial later when constructing a coordinate system for many colliding pointlike particles). Note also the special cases $p=\pm1$, when $\chi'$ is independent of $t'$, which just means that one of the surfaces bounding a wedge is a constant angle surface. A massless pointlike particle moving through \ads is shown in Figure \ref{1massless}, with $p=0$. The particle is annihilated at $t=\pi/2$ (to obtain a massless particle that bounces at the boundary, we should interpret the intersection of $w_+$ and $w_-$ as a massless particle also for $t>\pi/2$), and to 
see that the resulting spacetime is just empty \ads, we note that the resulting spacetime is just two halves of \ads, each written in boosted coordinate systems (with different boosts).

\section{The BTZ black hole}\label{secBTZ}
\begin{figure}
\includegraphics[scale=0.7]{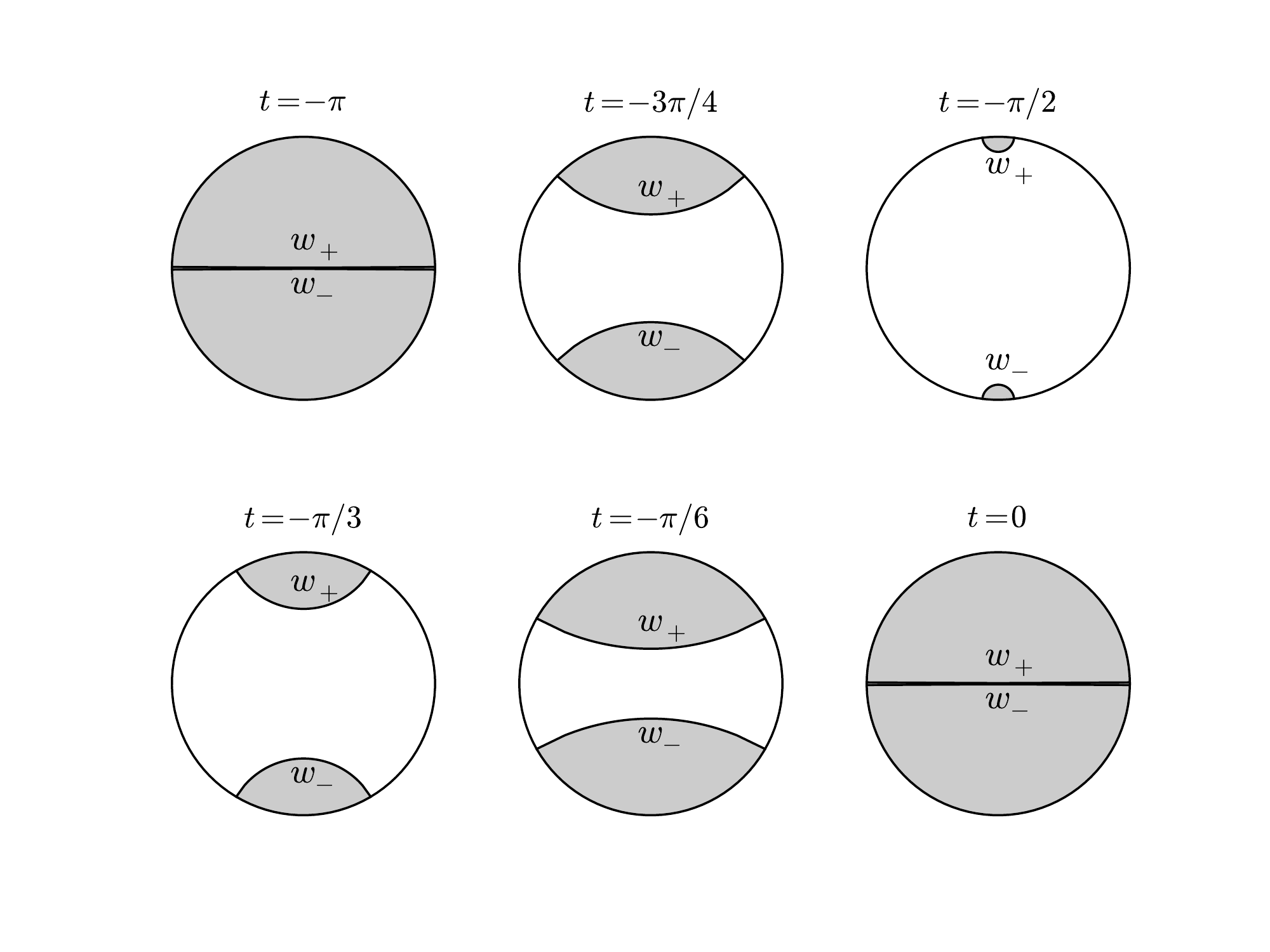}
\caption{\label{btzfig} The construction of the maximally extended BTZ black hole, by excising regions defined by equation \eqref{btzeq00}. The grey regions are removed parts of the spacetime, and the white region is the allowed part. $w_+$ and $w_-$ are identified, meaning that when we cross $w_\pm$ from the white region, we are mapped to $w_\mp$. In this example we have $\mu=2.8$.}
\end{figure}

\begin{figure}
\includegraphics[scale=0.7]{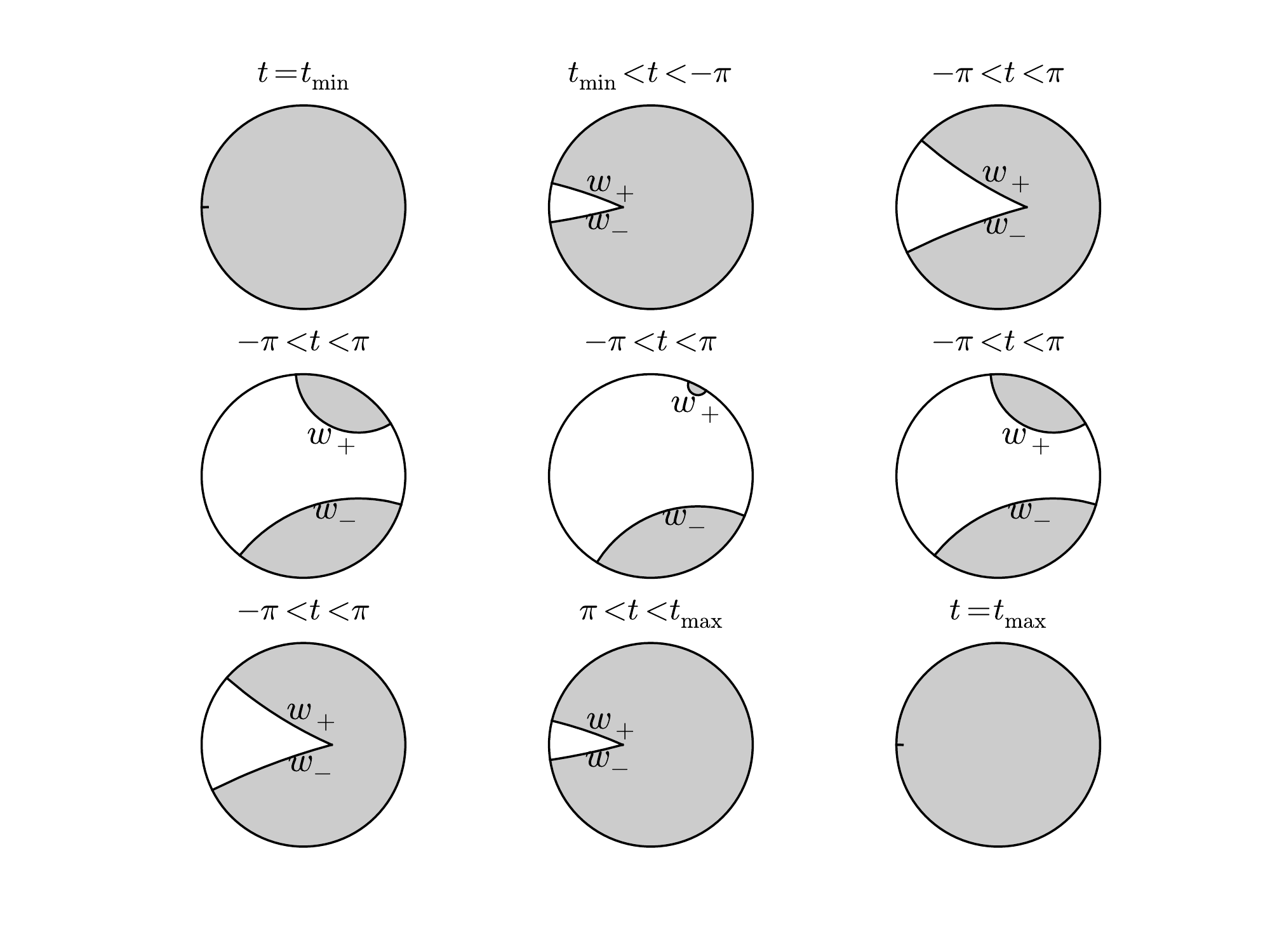}
\caption{\label{btzmodfig} The eternal BTZ black hole in the general coordinate system given by excising the regions defined by equation \eqref{btzeq}. $w_+$ and $w_-$ are identified, and the grey regions are the removed parts of the spacetime. The spacetime begins at $t=t_{\mathrm{min}}=-\arcsin(\tanh\zeta)-\pi$ and ends at $t=t_{\mathrm{max}}=\arcsin(\tanh\zeta)$. The parameters in this example are $\mu=1.8$, $\zeta=0.5$ and $\xi=1.1$ (but remember that $\mu$ is the only physically relevant parameter).}
\end{figure}

Just as a stationary massive particle can be described by identifying points under an isometry leaving the origin of \ads fixed (a rotation), the BTZ black hole can be obtained by identifying points under an isometry leaving a space-like geodesic fixed. We will for simplicity consider the spatial geodesic to lie along $\phi=0$ and $t=0$ and then in the end apply a trivial rotation to any angle. More precisely, we identify points under the isometry $u=e^{\mu\gamma_1}=\cosh\mu+\gamma_1\sinh\mu$. This leaves all geodesics of the form $\phi=0$ $t=n\pi$ invariant. Just like in the point particle case, we will look for curves of constant $t$ that are mapped to each other. If we assume that the curves are parametrized by
\begin{equation}
w_\pm=\cosh\chi\omega(t)+\sinh\chi\gamma(\pm\phi),
\end{equation}
which means that they are located symmetrically about the plane $\phi=0$. We can evaluate the equation $uw_-=w_+u$ to obtain the relation
\begin{equation}
\tanh\chi\sin \phi=-\sin t\tanh\mu,\label{btzeq00}
\end{equation}
where the curve $w_\pm$ has the coordinates $(\chi,t,\pm\phi)$. We will restrict $t$ to $-\pi\leq t \leq 0$. An illustration of this spacetime is shown in Figure \ref{btzfig}. However, for further purposes we will be interested in coordinates represented by wedges that are {\it not} located symmetrically around $\phi=0$. The easiest way to do this, is to apply yet another transformation characterized by the group element $e^{-\frac{1}{2}\xi\gamma_1}$. This transformation also has the spatial geodesic $(\phi=0,t=0)$ as fixed points. The lines $w_\pm$ transform into the new curves $w_\pm'=e^{\frac{1}{2}\xi\gamma_1}w_\pm e^{-\frac{1}{2}\xi\gamma_1}$ satisfying the relation
\begin{equation}
\tanh\chi\sin(\pm\phi)=-\sin t\tanh(\mu\pm\xi)\equiv-\sin t\tanh(\mu_\pm).\label{btzeq0}
\end{equation}
Points in this spacetime are still identified by the isometry $u=e^{\mu\gamma_1}$. What we have so far is a family of representations of a BTZ black hole where the singularity is represented by the spatial geodesic ($\phi=0,t=0$), and $\xi$ being the parameter parametrizing the family and $\mu$ is related to the mass of the black hole. Note that $\xi$ has no physical meaning and only specifies our coordinate system, analogous to the parameter $p$ for the pointlike particle. However, we would like to have a general parametrization of the black hole where the singularity is any radial spatial geodesic (analogous to the boosted pointlike particles). To this end we will do another transformation, which is a boost along $\phi=0$, represented by $u=e^{\frac{1}{2}\zeta\gamma_2}$. Using the general formulas \eqref{boosteqs}, we obtain that the lines are now given by
\begin{equation}
\tanh\chi(\cos\phi\sinh\zeta\sinh\mu_\pm\pm\sin\phi\cosh\mu_\pm)=-\sin t \cosh \zeta \sinh\mu_\pm
\end{equation}
Rotating such as to allow the geodesic to lie along an arbitrary angle $\psi$,  we can write this as
\begin{equation}
\tanh\chi\sin(-\phi+\psi+\Gamma_\pm)=-\sin t \coth \zeta \sin\Gamma_\pm,\label{btzeq}
\end{equation}
where
\begin{equation}
\tan \Gamma_\pm=\mp\tanh\mu_\pm\sinh\zeta.\label{Gammapmbh}
\end{equation}
This is very similar to the result of the boosted pointlike particle, except that we have $\coth\zeta$ instead of $\tanh\zeta$ in the right hand side, signifying the fact that the geodesic obtained by letting $\phi=\psi$ is space-like instead of time-like. An illustration of these coordinates is shown in Figure \ref{btzmodfig}.\\

\subsection{Transformation to Schwarzschild coordinates}\label{secSch}

A common metric for the non-rotating BTZ black hole is
\begin{equation}
\rd s^2=-\rd t^2 (-M+\rho^2/\ell^2)+\rd \rho^2\frac{1}{-M+\rho^2/\ell^2}+\rho^2 \rd \phi^2\label{BTZ},
\end{equation}
and we will call $M$ the mass of the black hole. An explicit way of seeing that the spacetime defined by \eqref{btzeq0} really is a BTZ black hole is to change coordinates to obtain the standard metric \eqref{BTZ}. The easiest way to do this is to realize that the parametrization \eqref{adscoord} of AdS is not optimal for understanding the effect of the isometry $e^{\mu\gamma_1}$, unlike in the case of the simple rotation used for the point particle, which in these coordinates just becomes a shift of the angular variable $\phi$. One may thus consider a different parametrization where we instead let 

\begin{equation}
x_0=\rho \cosh y,\hspace{10pt} x_2=\rho \sinh y. \label{BTZcoord}
\end{equation}
where $\rho\geq0$. The isometry $e^{\mu\gamma_1}: \bx \rightarrow e^{-\mu\gamma_1}\bx e^{\mu\gamma_1}$ is then just a translation $y\rightarrow y-2\mu$, and means that the curves in \eqref{btzeq0} correspond in these coordinates to surfaces of constant $y$. More precisely, from equation \eqref{btzeq0} we see that the surface $w_\pm$ corresponds to the surface with $y=\pm\mu_\pm=\xi\pm\mu$. However, the coordinates \eqref{BTZcoord} only parametrize the part where $|x_0|>|x_2|$, but one can note that the spacetime defined by \eqref{btzeq0} satisfies these bounds. We also have the restriction $x_0>0$, which is consistent with our choice of $-\pi\geq t\geq0$ for the time coordinate in the standard coordinates. The easiest way to parametrize the remaining coordinates is to assume that $|x_3|>|x_1|$ and let
\begin{equation}
x_1=\sqrt{1-\rho^2}\sinh t,\mathrm{  } x_3=\sqrt{1-\rho^2} \cosh t,
\end{equation}
which gives the metric \eqref{BTZ} (with $M=1$ and $y$ is identified with $\phi$) under the constraint $\rho<1$, which is the region inside the horizon. Assuming $|x_3|<|x_1|$ we have 
\begin{equation}
x_1=\sqrt{\rho^2-1} \cosh t,\mathrm{  } x_3=\sqrt{\rho^2-1} \sinh t,
\end{equation}
from which we also obtain the metric \eqref{BTZ}, but for $\rho>1$ which corresponds to the region outside the horizon. The $y$ coordinate plays the role of an angular variable and becomes compactified under the identification of the isometry, which in these coordinates reads $y\sim y-2\mu$. To have the standard range $2\pi$ for the angular variable, we may rescale $y=\mu \phi /\pi$,$\rho\rightarrow\rho\pi/\mu$, $t\rightarrow t\mu/\pi$, from which we obtain the metric \eqref{BTZ} with $M=\mu^2/\pi^2$.

\subsection{Merging many black holes}\label{secMerge}
Just as a pointlike particle can be represented by $n$ wedges, where the border of one wedge is identified with the next border of the next one, a BTZ black hole can be represented by $n$ copies of spacetimes defined by equation \eqref{btzeq0}, that are mapped to each other, forming a BTZ black hole with a larger mass. In such a construction, we would have $n$ spacetimes defined by \eqref{btzeq}, with parameters $(\Gamma_\pm^i,\zeta_i)$, and where $w_+^i$ is mapped to $w_-^{i+1}$. As shown previously, each such spacetime can be written as the metric
\begin{equation}
\rd s^2=-\rd t^2 (-1+\rho^2/\ell^2)+\rd \rho^2\frac{1}{-1+\rho^2/\ell^2}+\rho^2 \rd y^2,
\end{equation}
such that the $w_\pm^i$ will be located at $y=\pm\mu_i$. We thus glue all these spacetimes together along the angular coordinate $y$, such that the total periodicity is $2\sum_i\mu_i$, and thus the total mass of the black hole is $M=(\sum_i\mu_i)^2/\pi^2$.

\section{Collision of particles}\label{collisions}
In this section we will explain how to construct a solution corresponding to colliding massless pointlike particles. We will be concerned only with particles falling on radial geodesics, such that all geodesics intersect at exactly one point $(t=0,\chi=0)$, and we will assume that when the particles collide they form a joint object. We will first briefly review the case of two particles in a head-on collision with equal energies, which was studied in \cite{Matschull:1998rv} and we refer the reader to that paper for more details.

\subsection{Collision of two particles}
\begin{figure}
\includegraphics[scale=0.7]{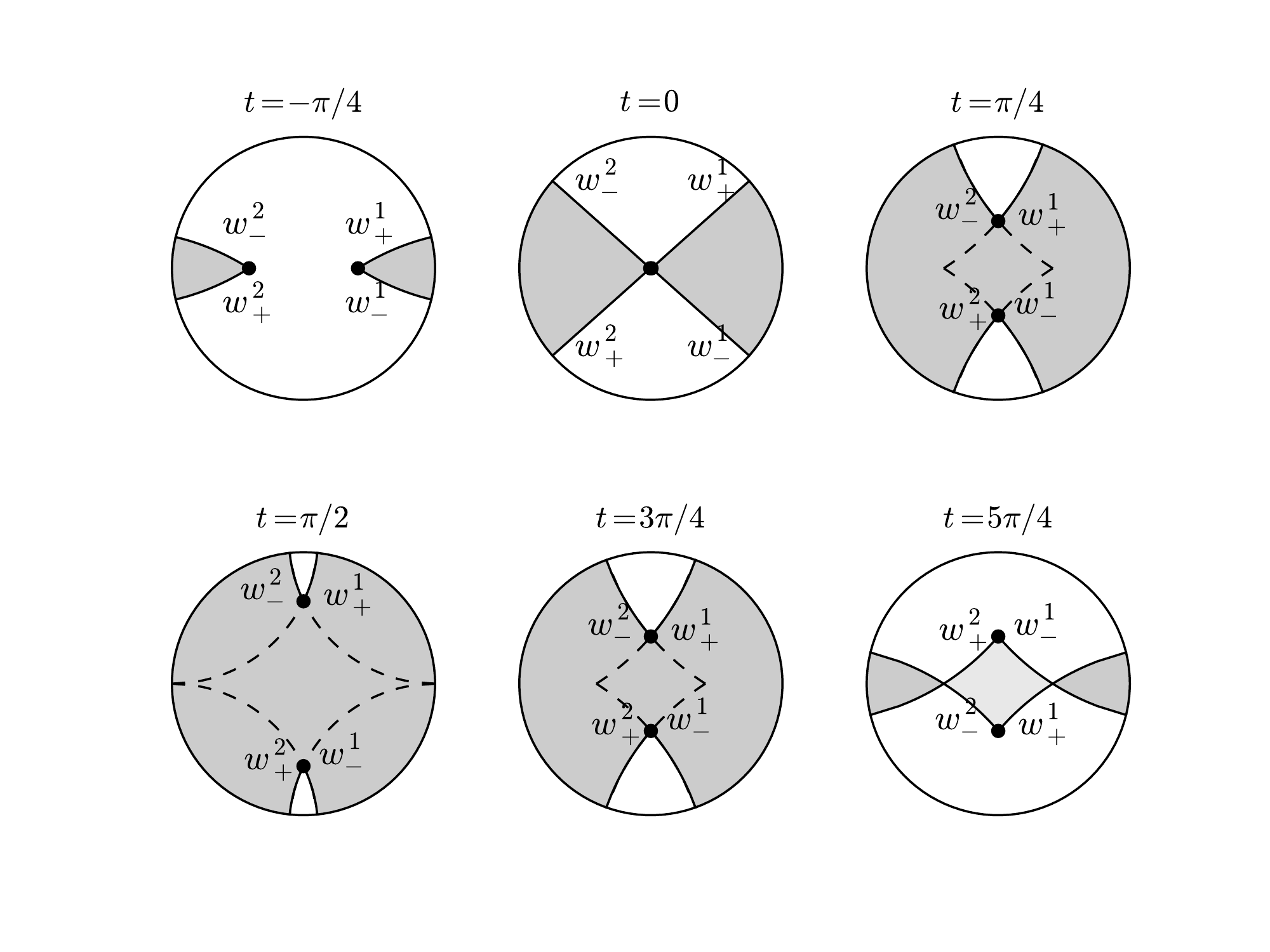}
\caption{\label{2particles_pp} Collision of two particles, with equal energy, falling radially on geodesics at angles $\psi_i=(0,\pi)$ and with $e_i=(0.9,0.9)$. The dark grey areas constitute removed pieces. The resulting object moves on a time-like geodesic, so a massive point particle has been formed. The light grey regions in the last panel marks that we have two allowed patches on top of each other. The two oscillating geodesics which are identified and represent the resulting massive particle, will keep oscillate forever in this coordinate system. The dashed lines represent the wedges for each of the two particles if the there was only one particle and thus no collision.}
\end{figure}

\begin{figure}
\includegraphics[scale=0.7]{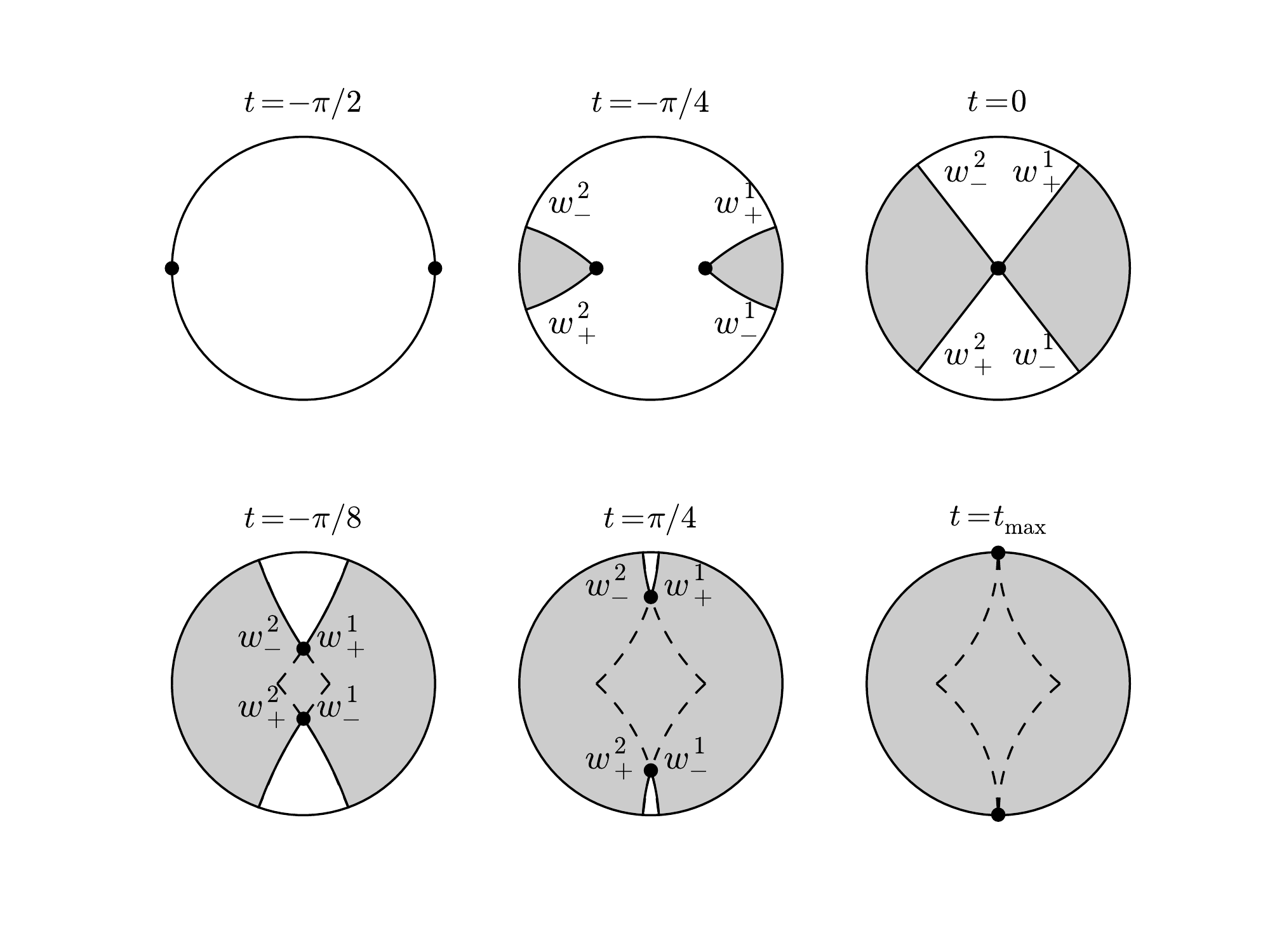}
\caption{\label{2particles_bh} Collision of two particles, with equal energy, falling radially on geodesics at angles $\psi_i=(0,\pi)$ and with $e_i=(1.3,1.3)$. The dark grey areas constitute removed pieces. The resulting object moves on a space-like geodesic, so a BTZ black hole has formed. The spacetime in these coordinates ends at $t=t_{\mathrm{max}}\approx0.88$, but this is just a coordinate artefact. If we would transform the late time geometry to the more conventional metric \eqref{BTZ} for the BTZ black hole, the time coordinate in these coordinate would go to infinity when $t\rightarrow t_{\mathrm{max}}$.}
\end{figure}

In this section we will review the two-particle case, and explain why this construction does not trivially generalize to more particles. We will assume that the two particles have the same energy and are colliding head on at the point $(t=0,\chi=0)$ (we can always do a coordinate transformation to achieve this). The two particles are then constructed by excising a wedge described by equation \eqref{masslessp} behind each particle. Due to the symmetries of the problem, we can assume that $p_1=p_2=0$, where $p_1$ and $p_2$ are the parameters specifying the wedges in equation \eqref{Gammapmmassless}, for particle 1 and 2 respectively. The part of space that is removed behind particle 1 respectively particle 2 are thus bounded by two surfaces $w_\pm^1$ respectively $w_\pm^2$. This means that before the collision ($t<0$), particle 1 is described by cutting out a wedge as in Figure \ref{1massless}, while particle 2 is the same but rotated 180 degrees. The holonomies are thus given by
\begin{equation}
\bu_1=1+\tan\epsilon (\gamma_0-\gamma_1),
\end{equation}
for particle 1 and 
\begin{equation}
\bu_2=1+\tan\epsilon (\gamma_0+\gamma_1),
\end{equation}
for particle 2. In these coordinates there is a very natural way to construct the spacetime corresponding to the formation of a joint object. We just identify the joint object with the intersection between $w_+^1$ and $w_-^2$ (which is identified with the intersection between $w_+^2$ and $w_-^1$ via the holonomies of the particles). We easily obtain from equation \eqref{masslessp} that the resulting geodesic is given by $\tanh\chi=\tan\epsilon\sin t$, moving on angles $\psi=\pm\pi/2$. This is a timelike geodesic if $|\tan\epsilon|<1$, and spacelike if $|\tan\epsilon|>1$, and will be identified with a massive pointlike particle in the former case and a BTZ black hole in the latter. This condition can also obtained by computing the holonomy of the resulting object, which is given by the product of the holonomies of the two particles, as was done in \cite{Matschull:1998rv}. In Figure \ref{2particles_pp} the formation of a massive pointlike particle is shown, and Figure \ref{2particles_bh} illustrates the 
formation of a BTZ black hole. \\

To see that the resulting spacetime really is that of a BTZ black hole or a massive pointlike particle, we would like to identify the resulting geometry using either equation \eqref{c1} or \eqref{btzeq}. We can do this by rewriting the equations governing $w_\pm^i$. As can be seen in Figure \ref{2particles_pp} or \ref{2particles_bh}, after the collision the surfaces $w_+^1$ and $w_-^2$ will bound a new wedge of allowed geometry, that is connected to a geodesic moving along an angle $\psi=\pi/2$. To make this manifest, we rewrite them in the following way\\
\begin{equation}
w_+^1: \tanh\chi'\sin(-\phi'+\epsilon)=-\sin\epsilon\sin t \Rightarrow \tanh\chi'\sin(-\phi'+\frac{\pi}{2}+(\epsilon-\frac{\pi}{2}))=\tan\epsilon \sin(\epsilon-\frac{\pi}{2})\sin t',
\end{equation}
\begin{equation}
w_-^2: \tanh\chi'\sin(-\phi'-\epsilon+\pi)=\sin\epsilon\sin t \Rightarrow \tanh\chi'\sin(-\phi'+\frac{\pi}{2}+(\frac{\pi}{2}-\epsilon))=\tan\epsilon \sin(\frac{\pi}{2}-\epsilon)\sin t'.
\end{equation}
When the resulting object is a massive point particle, we can compare this with equation \eqref{c1} to obtain that this is a wedge with $\Gamma_\pm=\pm(\frac{\pi}{2}-\epsilon)$ and $\tanh \zeta=-\tan\epsilon$ (note that $\zeta$ is negative, but this is not a problem and could be fixed by just changing the angle of the geodesic by $\pi$). In the coordinate system where this wedge is stationary, we thus obtain from \eqref{Gammapm} that it is a circle segment with a total angle $2\nu$, where $\nu$ is given by 
\begin{equation}
\tan(\pi/2-\epsilon)=\tan\nu\cosh\zeta \Rightarrow \cos\nu=\tan\epsilon.
\end{equation}
The wedge bounded by $w_+^2$ and $w_-^1$ works in exactly the same way, so the total deficit angle is $2\pi-4\nu$, and the mass of the particle is thus $m=\pi-2\nu$, and using the above relation we obtain $\sin(m/2)=\tan\epsilon$, which agrees with the result in \cite{Matschull:1998rv}, which was obtained by just taking the product of the two holonomies. The method carried out here might seem more cumbersome, but it will be useful for the many particle case.\\

If $|\tan\epsilon|>1$, we can not identify this with \eqref{c1} anymore, but instead we identify this to be a part of a BTZ black hole, and match the parameters to equation \eqref{btzeq} and \eqref{Gammapmbh}. We thus obtain $\coth\zeta=-\tan\epsilon$, and we still have $\Gamma_\pm=\pm(\frac{\pi}{2}-\epsilon)$. From \eqref{Gammapmbh} we now have
\begin{equation}
\tan\epsilon=-\tanh\mu\sinh\zeta\Rightarrow \cosh\mu=\tan\epsilon
\end{equation}
Since we have again two wedges, the total mass parameter $M$ can then be obtained as $M=4\mu^2/\pi^2$.\\

In this section the wedges corresponding to each particle were located symmetrically behind the particle. Will this also work for other setups with more particles? The answer is generically no, and this can be seen already in the two-particle case. Let's assume for example that the two particles are not moving on exactly opposite angles, say $\psi_1=0$ but $\psi_2\neq\pi$. We could then try to play the same game, excising a symmetric wedge behind each particle, and then try to identify the intersections as a joint object. The problem here is that, even though $w_+^i$ is mapped to $w_-^i$ by $\bu_i$, the intersection between $w_+^1$ and $w_-^2$ will generically not be mapped to the intersection between $w_-^1$ and $w_+^2$, so we can not identify this as a joint object and this coordinate system breaks down after the collision. To solve this problem, one must allow for more general wedges, specified by parameters $p_1$ and $p_2$ acoording to equation \eqref{masslessp}, and then adjust these parameters such 
that the intersections are mapped to each other. For the two-particle case, this is not necessary since we can always make a coordinate transformation such that the particles have equal energies and are colliding head on such that the wedges are symmetrical. For many particles this is not the case, and it is crucial to allow for more general wedges to be able to construct consistent solutions for arbitrary initial conditions. We will explore this in more detail in the next section.

\subsection{Collision of $n$ particles}
We will in this section consider the construction of a solution corresponding to $n$ colliding massless pointlike particles, falling in along radial geodesics at arbitrary angles and with arbitrary energies. Each particle is described by excising a wedge behind it borderd by two surfaces $w_\pm^i$ governed by equation \eqref{masslessp}, with parameters $\psi_i$, $e_i=\tan\epsilon_i$ and $p_i$, for $i=1,\ldots,n$ and we identify $n+1$ with $1$. We thus assume that the particles are all created at time $t=-\pi/2$, such that they all collide at $t=0$. $p_i$ has no physical meaning and just parametrizes our different coordinate systems.

\begin{figure}
\includegraphics[scale=0.7]{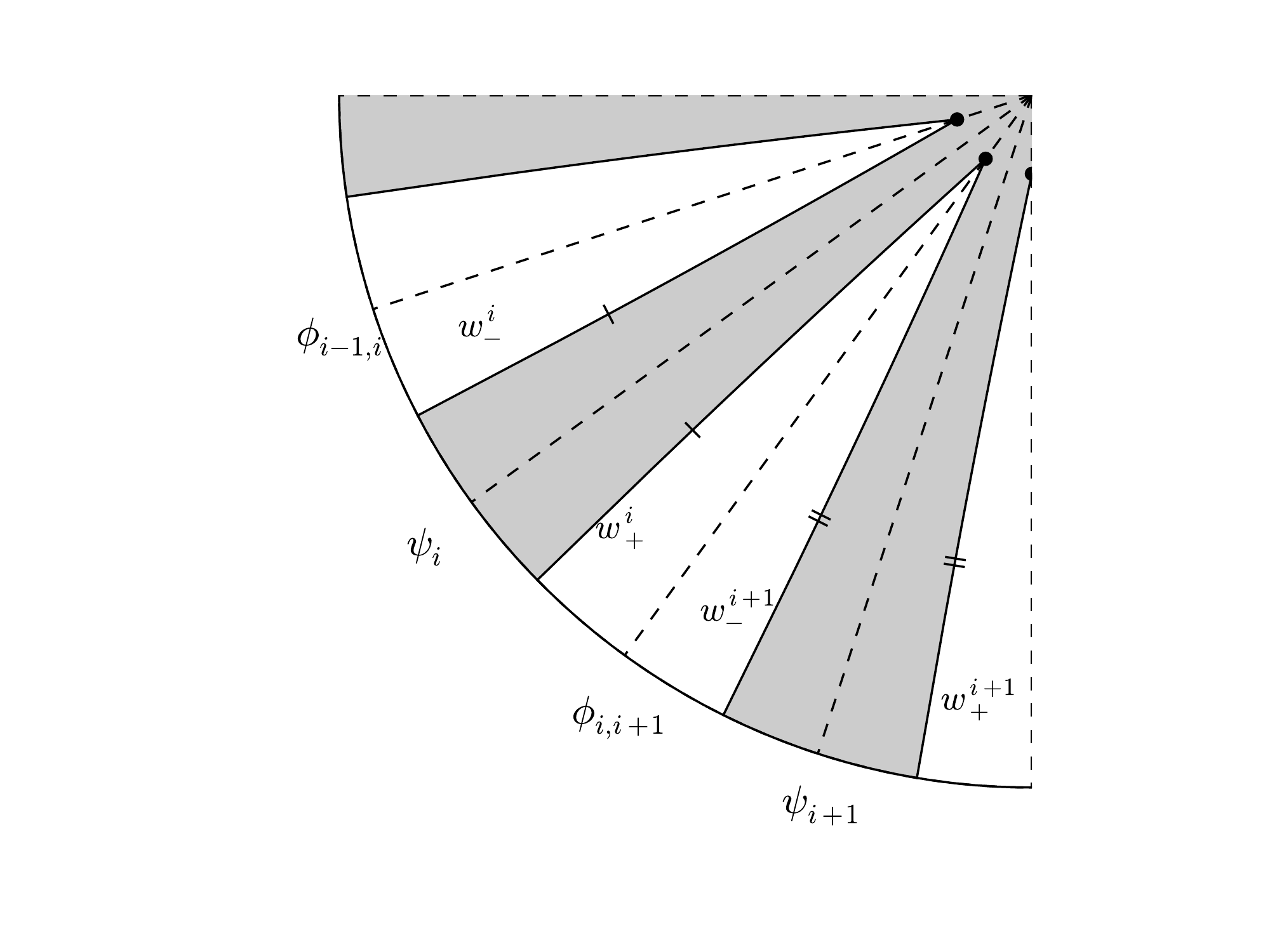}
\caption{\label{illustration} An illustration of the different parameters after the collision has occurred. The grey regions are the removed parts of the spacetime. The particles fall in along the angles $\psi_i$, and the wedge associated to particle $i$ is bounded by $w_\pm^i$. The joint object then moves on angles $\phi_{i,i+1}$, which is the intersection of $w_+^i$ and $w_-^{i+1}$. Parameters with index $i$ (for example $\Gamma_\pm^i$, $p_i$ and $e_i$) are associated to the (grey) wedge at angle $\psi_i$, while parameters with index $i,i+1$ (for example $\Gamma_\pm^{i,i+1}$, $\zeta_{i,i+1}$ and $\nu_{i,i+1}$) are associated to the (white) wedge at angle $\phi_{i,i+1}$.}
\end{figure}

After the collision we will identify the intersections of $w_+^i$ and $w_-^{i+1}$ with the resulting joint object. For this to be consistent, we must have that the intersection between $w_+^i$ and $w_-^{i+1}$ must be mapped by the isometry $u_{i}$ corresponding to particle $i$ to the intersection between $w_+^{i-1}$ and $w_-^{i}$. As we will see we can choose the parameters $p_i$ in such a way to accomplish this. The case of $n$ particles with identical energies, evenly distributed along the angular direction (for which the two-particle setup considered here is a special case), is trivial in the sense that, by symmetry, we can choose wedges that are located symmetrically behind each particle, meaning $p_i=0$. However, if the particles are located at arbitrary angles, or with arbitrary energies, this is not possible. An illustration of various variables discussed in this section is shown in Figure \ref{illustration}.\\

\begin{figure}
\includegraphics[scale=0.7]{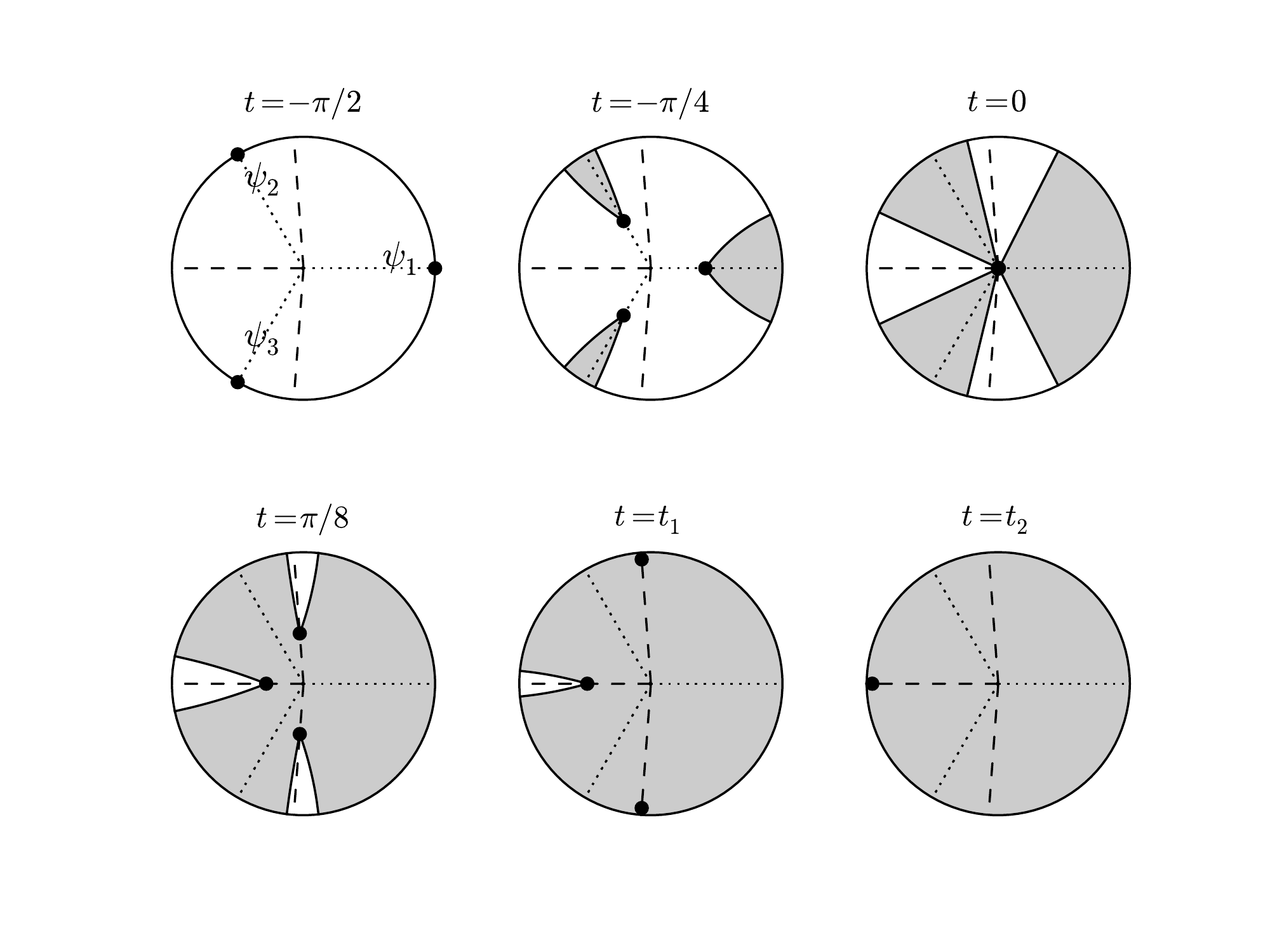}
\caption{\label{3particles} Collision of three particles, falling radially on geodesics at angles $\psi_i=(0,2\pi/3,4\pi/3)$ and with $e_i=(2,\frac{1}{2},\frac{1}{2})$. The grey areas constitute removed pieces. The resulting object moves on spatial geodesics (marked by dashed lines), so a black hole has formed instead of a massive particle. The dotted lines are the trajectories of the massless particles, and we see clearly that the wedges for particle 2 and 3 are not located symmetrically behind the particles. Note that lines corresponding to one grey wedge are not nessecarily mapped to each other on the same time slice, which is why it is possible for some of the white wedges in the last panels to disappear first at time $t_1\approx0.61$ (essentially, when jumping from one white area to another when going around one particle, there might be a time shift, which would not be present for wedges that are located symmetrically behind the particle as in the two-particle case). The last wedge disappears at time $t_
2\approx0.82$, but these two times are of course mapped to each other by the isometries of the particles. The parameters specifying the wedges, obtained by numerically 
solving \eqref{peq} and \eqref{intersection}, are $p_i\approx(0, 0.409, -0.409)$, and the angles of the intersections are $\phi_{i,i+1}\approx(1.644,  \pi, -1.644)$.}
\end{figure}

\begin{figure}
\includegraphics[scale=0.7]{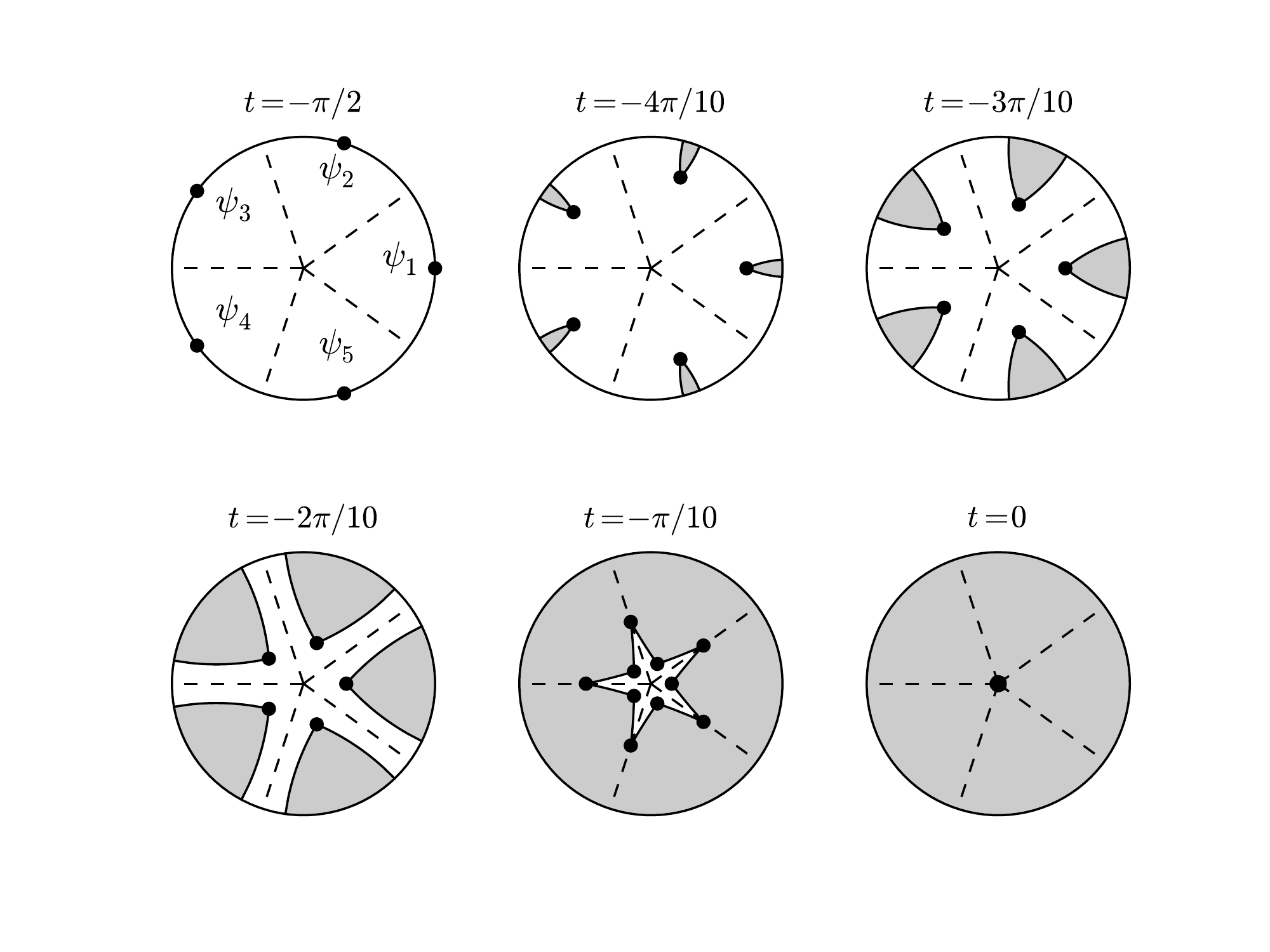}
\caption{\label{5particles} Collision of five particles, with equal energy, falling radially on geodesics at angles $\psi_i=(0,2\pi/5,4\pi/5,6\pi/5,8\pi/5)$ and with $e_i=1.4$. The grey areas constitute removed pieces. The resulting joint object moves on spatial geodesics (marked with dashed lines), and thus a black hole has formed. Note that in this case the resulting spatial geodesics go back in time in these coordinates, but this is purely a coordinate artifact and has no physical meaning.}
\end{figure}

The intersection of the surface $w_+^i$ belonging to wedge $i$ and $w_-^{i+1}$ belonging to wedge $i+1$ will be on a line of constant angle $\phi_{i,i+1}$ (a radial geodesic), and is obtained from \eqref{masslessp} as the solution of the equation 
\begin{equation}
\frac{\sin(\phi_{i,i+1}-\Gamma_i^+-\psi_i)}{\sin(\phi_{i,i+1}-\Gamma_{i+1}^--\psi_{i+1})}=\frac{\sin\Gamma_i^+}{\sin\Gamma_{i+1}^-}.\label{intersection0}
\end{equation}
After the collision, we identify the new object with the intersections of the surfaces bounding two neighboring wedges. This is only consistent if these intersections are mapped to each other by the isometries associated to the particles. So let the intersection between line $w_+^i$ and $w_-^{i+1}$ be denoted by $I_{i,i+1}$ and let $u_i$ be the holonomy of particle $i$. We then must have the condition
\begin{equation}
I_{i-1,i}=u_i^{-1}I_{i,i+1}u_i,
\end{equation}
for all $i$. This gives us $n$ conditions to fix the $n$ parameters $p_i$. The holonomies are here given by
\begin{equation}
\bu_i=1+e_i(\gamma_0-\gamma(\psi_i)).
\end{equation}
Assume now that we know the angles $\phi_{i-1,i}$ for intersection $I_{i-1,i}$ and $\phi_{i,i+1}$ for intersection $I_{i,i+1}$, and let us see how we can use this to fix the parameter $p_i$ for particle $i$. Since $I_{i-1,i}$ is the intersection between the plane with constant angle $\phi_{i-1,i}$ and the surface $w_-^i$, they satisfy the equations
\begin{equation}
\tanh\chi'\sin(-\phi_{i-1,i}+\Gamma_-^i+\psi_i)=-\sin\Gamma_-^i\sin t',
\end{equation}
where
\begin{equation}
\tan\Gamma_-^i=-(1-p_i)e_i,
\end{equation}
and since $I_{i,i+1}$ is the intersection between $\phi_{i,i+1}$ and $w_+^i$, we have the equations
\begin{equation}
\tanh\chi\sin(-\phi_{i,i+1}+\Gamma_+^i+\psi_i)=-\sin\Gamma_+^i\sin t,
\end{equation}
where
\begin{equation}
\tan\Gamma_+^i=(1+p_i)e_i.
\end{equation}
Our consistency condition now forces us to make sure that these two lines must be mapped to each other by the holonomy $\bu_i=1+e_i(\gamma_0-\gamma(\psi_i))$, and by evaluating this (see Appendix \ref{app2}) we obtain
\begin{equation}
p_i=\frac{\tan(\phi_{i,i+1}-\psi_i)+\tan(\phi_{i-1,i}-\psi_i)}{-2e_i\tan(\phi_{i,i+1}-\psi_i)\tan(\phi_{i-1,i}-\psi_i)+\tan(\phi_{i,i+1}-\psi_i)-\tan(\phi_{i-1,i}-\psi_i)}.\label{peq}
\end{equation}
Equation \eqref{intersection0} can be written as
\begin{equation}
-\sin(\phi_{i,i+1}-\psi_i)\frac{1}{e_i(1+p_i)}+\cos(\phi_{i,i+1}-\psi_i)=\sin(\phi_{i,i+1}-\psi_{i+1})\frac{1}{e_{i+1}(1-p_{i+1})}+\cos(\phi_{i,i+1}-\psi_{i+1})\label{intersection}.
\end{equation}
Equation \eqref{peq} and equation \eqref{intersection} constitute $2n$ equations to solve for the $2n$ variables $\phi_{i,i+1}$ and $p_i$, for $i=1,\ldots,n$. Solving this will give a consistent slicing of the spacetime which can be continued after the collision. However, except for the case of discrete rotational symmetry (for which $p_i=0$), it seems difficult to find an analytic solution to these equations. We can solve this numerically and we see as expected that there is a solution that is continuously connected to the rotationally symmetric case (there might be other solutions, but we will not investigate this question in this paper). In Figure \ref{3particles} we illustrate this with the example of three particles, located at angles $\psi_i=(0,2\pi/3,4\pi/3)$ and with energies determined by $e_i=(2,\frac{1}{2},\frac{1}{2})$. Numerically solving \eqref{peq} and \eqref{intersection} we obtain $p_i\approx(0, 0.409, -0.409)$ and $\phi_{i,i+1}\approx(1.644,  \pi, -1.644)$. In this figure it is clear that 
the wedges are not located symmetrically about the particles. Note also that the surfaces $w_\pm^i$ are not mapped to each other within constant time slices. This explains why it seems like space disappears earlier in two of the wedges, because this is just a coordinate artefact and the time $t_1$ (when the first two wedges disappear) is mapped to time $t_2$ (when the last wedge disappears) by the holomies of particles 2 and 3. In Figure \ref{5particles} we show an example of collision of five particles in the case of discrete rotational symmetry (so the wedges can be chose symmetrically behind each particle). Here the resulting object moves on spatial geodesics that go back in (coordinate) time, but this has no physical significance and is also just a coordinate artifact.

\section{Post collision geometry}\label{geometry}
To investigate the geometry after the collision, we would like to make a coordinate transformation of the late time geometry such that it is either manifestly a massive pointlike particle or a black hole. Just as in the two-particle case, this can be done by mapping the resulting wedges of allowed geometry (for instance the white parts in the last panels in Figure \ref{3particles}), to either the equations describing a boosted conical singularity (as described by equation \eqref{c1}), or a boosted BTZ black hole (as described by equation \eqref{btzeq}). We thus have a set of new wedges of allowed geometry, and we will associate the wedge bounded by $w_+^i$ and $w_-^{i+1}$ with an index $(i,i+1)$, which thus moves along an angle $\psi=\phi_{i,i+1}$. We will now rewrite the equations governing $w_+^i\equiv w_-^{i,i+1}$ and $w_-^{i+1}\equiv w_+^{i,i+1}$ to make it manifest that they satisfy either equation \eqref{c1} or \eqref{btzeq} for some new parameters $\Gamma_\pm^{i,i+1}$ and $\zeta_{i,i+1}$. For $w_+^i$ 
we have the equation
\begin{align}
\tanh\chi'\sin(-\phi'+\psi_i+\Gamma_+^i)&=-\sin \Gamma_+^i\sin t'\nonumber
\\
&\Leftrightarrow\nonumber
\\
\tanh\chi'\sin(-\phi'+(\psi_i+\Gamma_+^i-\phi_{i,i+1})+\phi_{i,i+1})&=-\frac{\sin \Gamma_+^i}{\sin(\psi_i+\Gamma_+^i-\phi_{i,i+1})}\nonumber\\
&\times\sin(\psi_i+\Gamma_+^i-\phi_{i,i+1})\sin t',\label{rec1}
\end{align}
and for $w_-^{i+1}$ we have
\begin{align}
\tanh\chi'\sin(-\phi'+\psi_{i+1}+\Gamma_-^{i+1})&=-\sin \Gamma_-^{i+1}\sin t'\nonumber
\\
&\Leftrightarrow\nonumber
\\
\tanh\chi'\sin(-\phi'+(\psi_{i+1}+\Gamma_-^{i+1}-\phi_{i,i+1})+\phi_{i,i+1})&=-\frac{\sin \Gamma_-^{i+1}}{\sin(\psi_{i+1}+\Gamma_-^{i+1}-\phi_{i,i+1})}\nonumber\\
&\times\sin(\psi_{i+1}+\Gamma_-^{i+1}-\phi_{i,i+1})\sin t'.\label{rec2}
\end{align}
Now the definition of $\phi_{i,i+1}$ as the intersection of $w_+^i$ and $w_-^{i+1}$ requires that
\begin{equation}
\frac{\sin \Gamma_-^{i+1}}{\sin(\psi_{i+1}+\Gamma_-^{i+1}-\phi_{i,i+1})}=\frac{\sin \Gamma_+^i}{\sin(\psi_i+\Gamma_+^i-\phi_{i,i+1})}\equiv \Bigg\{\begin{array}{ll}
                                                                                                                                             \tanh\zeta_{i,i+1},&\text{Point particle}\\
                                                                                                                                             \coth\zeta_{i,i+1},&\text{Black hole}                                                                                                                                             
                                                                                                                                            \end{array}\label{zetadef}
\end{equation}
depending on if the magnitude of this quantity is smaller or larger than one. We thus see that, by comparing to either \eqref{c1} or \eqref{btzeq}, that the patch enclosed by these surfaces can indeed be interpreted as a boosted wedge bounded by surfaces $w_\pm^{i,i+1}$ satisfying the equations
\begin{equation}
\tanh\chi'\sin(-\phi'+\Gamma_\pm^{i,i+1}+\phi_{i,i+1})=-\kappa\sin\Gamma_\pm^{i,i+1}\sin t',
\end{equation}
with 
\begin{align}
\Gamma_-^{i,i+1}&=\psi_i+\Gamma_+^i-\phi_{i,i+1},\nonumber\\
\Gamma_+^{i,i+1}&=\psi_{i+1}+\Gamma_-^{i+1}-\phi_{i,i+1}.\label{Gammaiis}
\end{align}
and
\begin{equation}
\kappa=\Bigg\{\begin{array}{ll}
\tanh\zeta_{i,i+1},&\text{Point particle}\\
\coth\zeta_{i,i+1},&\text{Black hole}                                                                                                                                             
\end{array}
\end{equation}

For the pointlike particle case, in the coordinate system where this particle is stationary, this patch is a circle segment. From equation \eqref{Gammapm} we can then read off that the parameters $p_{i,i+1}$ and $\nu_{i,i+1}$ associated to this circle segment are given implicitly by the equations
\begin{equation}
\tan \Gamma_\pm^{i,i+1}=\pm\tan\left((1\pm p_{i,i+1})\nu_{i,i+1}\right)\cosh\zeta_{i,i+1},\label{Gammaii}
\end{equation}
such that the angle of this segment is $2\nu_{i,i+1}$. The total deficit angle is then $2\pi-2\sum_i \nu_{i,i+1}$ (recall that these segments now define the {\it allowed} geometry, and not the {\it removed} geometry, which is the reason for why the deficit angle is not just $2\sum_i\nu_{i,i+1}$).\\

For the black hole we want to relate these parameters to the black hole mass and $\mu_{i,i+1}$, and we find that
\begin{equation}
\tan\Gamma_\pm^{i,i+1}=\mp\tanh\mu^{i,i+1}_\pm\sinh\zeta_{i,i+1},\label{Gammaiibh}
\end{equation}
where $\mu_\pm^{i,i+1}\equiv\mu_{i,i+1}\pm\xi_{i,i+1}$. The total mass of the black hole is simply
\begin{equation}
M=\frac{1}{\pi^2}\left[\sum_i\mu_{i,i+1}\right]^2.
\end{equation}
Note that under the isometries $u_i$ corresponding to the massless particles, space-like (time-like) geodesics are mapped to space-like (time-like) geodesics. This is why we know that (the absolute value of) equation \eqref{zetadef} is either larger than one for all $i$, or smaller than one for all $i$ (or equal to one in the extremal case, but we will not consider that in this paper). This is built in into the construction of the consistency equations \eqref{peq} and \eqref{intersection}, although it seems difficult to prove it directly from these equations. \\

Notice that the coordinate transformations to bring the metric to the static metric for a conical singularity or the Schwarschild metric for the black hole, will map the light-like geodesics that pass through the origin to new light-like geodesics passing through the origin. Thus, if we let \LS denote the light-like surface corresponding to the set of all light-like geodesics that end at the origin and originate at $t'=-\pi/2$, we can do the coordinate transformation to the static coordinate system on the patch {\it above} the surface \LS. The resulting spacetime will then be the static conical singularity metric, or the Schwarschild AdS black hole metric, glued to empty AdS across a light-like surface, and where the massless point particles are moving inside this surface. We will explore this further in the next section when we investigate the $n\rightarrow\infty$ limit.

\section{The $n\rightarrow\infty$ limit}\label{secLimit}

In this section we will consider what happens to the equations \eqref{peq} and \eqref{intersection} when we take the number of particles going to infinity. For $n$ particles, the setup is described by $n$ triples $(e_i,\psi_i,p_i)$. We will assume that in the limit, for every $\phi$ and for every $\delta\phi$, we can find sufficiently large $n$ such that $(\phi,\phi+\delta\phi)$ contains arbitrarily many $\psi_i$. In such a limit, the setup is expected to degenerate into only a continuous energy density function $\rho(\phi)$. For simplicity we will assume that $\psi_i=2\pi i/n$, such that $\psi_{i-1}-\psi_i\equiv\rd\phi=2\pi/n$, although the result will not depend on the precise way the limit is taken. For $n>>1$, we expect that we can approximate the $e_i$ and $p_i$ by continuous functions. Thus, to obtain the correct limit for the energy density, we define $e_i=\rho(\psi_i)\rd\phi$. It turns out that for large $n$ we have $p_i\sim n$ and that $\phi_{i,i+1}-\psi_i$ approaches a constant (which is 
generically not zero). We will thus define continuous interpolating functions $P(\phi)$ and $\Phi(\phi)$ such that $e_ip_i=P(\psi_i)$ and $\phi_{i,i+1}-\psi_i=\Phi(\psi_i)$. The equations \eqref{peq} and \eqref{intersection} will in the $n\rightarrow\infty$ limit become differential equations for $P$ and $\Phi$, with $\rho$ as a source. Using $P(\psi_{i+1})=P(\psi_i)+P'(\psi_i)\rd\phi+O(1/n^2)$ and $\Phi(\psi_{i+1})=\Phi(\psi_i)+\Phi'(\psi_i)\rd\phi+O(1/n^2)$, it is straightforward to deduce from \eqref{peq} and \eqref{intersection}, that in the limit $n\rightarrow\infty$, $P$ and $\Phi$ must satisfy the differential equations
\begin{subequations}
\begin{align}
P'\tan \Phi&=(2\rho-P^2)\tan\Phi-P\label{ode1},\\
P(\tan\Phi)'&=2\rho\tan\Phi+(2\rho-1)P\tan^2\Phi -P,\label{ode2}
\end{align}
\end{subequations}
with periodic boundary conditions. However, when deriving the above equations, we had to use that $|e_i|<<|p_ie_i|$, which is in general valid when $n\rightarrow\infty$, but it is not valid if $p_i=0$ exactly. Or in other words, the derivation of \eqref{ode1} and \eqref{ode2} is not valid in the rotationally symmetric case, since the limits $n\rightarrow\infty$ and the limit $\rho\rightarrow$const. do not commute. In the rotationally symmetric case we instead have the solution $P=0$ and $\Phi=0$. This seems to be a solution of \eqref{ode1} and \eqref{ode2} for any $\rho$, but this is also an artifact of the fact that the derivation of these equations is problematic in this case (of course it is not a problem if $P$ or $\Phi$ vanish at isolated points which generically will happen, it is only problematic if $P\equiv0$ on a continuous interval which seems to only happen in the case of rotational symmetry). In practice, it turns out to be easier to find the correct solutions of \eqref{ode1} and \eqref{ode2} 
by 
directly solving \eqref{intersection} and \eqref{peq} for some large $n$, but one can then verify that these solutions indeed satisfy \eqref{ode1} and \eqref{ode2} up to $O(1/n)$.\\

For further purposes we will need $\Gamma_\pm^{i,i+1}$ expressed in terms of $P$ and $\Phi$. From \eqref{Gammaiis} we obtain
\begin{equation}
\tan\Gamma_-^{i,i+1}=\frac{P-\tan\Phi}{1+\tan\Phi P}+\left[\frac{1}{1+P\tan\Phi}-\frac{P-\tan\Phi}{(1+\tan\Phi P)^2}\tan\Phi\right]\rho \rd\phi+O(\frac{1}{n^2}),
\end{equation}
\begin{align}
\tan\Gamma_+^{i,i+1}&=\frac{P-\tan\Phi}{1+\tan\Phi P}+\Bigg[\frac{-\rho+P'+1+\tan^2\Phi}{1+P\tan\Phi}-\nonumber\\
-&\frac{(P-\tan\Phi)((P'-\rho)\tan\Phi-P-P\tan^2\Phi)}{(1+\tan\Phi P)^2}\Bigg]\rd\phi+O(\frac{1}{n^2}),
\end{align}
and thus
\begin{align}
\tan\Gamma_+^{i,i+1}-\tan\Gamma_-^{i,i+1}=\frac{1}{(1+P\tan\Phi)^2}(1+\tan^2\Phi)(-2\rho+P'+1+P^2)\rd\phi.\label{tandiff}
\end{align}

\subsection{Conical singularity}
In the case of the formation of a conical singularity (massive pointlike particle), the resulting spacetime will in the limit be that of a conical singularity geometry and empty AdS glued together across a light-like surface. Let us denote this surface \LS, which thus are all points which satisfy $\tanh\chi'=-\sin t'$, $-\pi/2\leq t \leq 0$. Above the shell we have the geometry of a massive pointlike particle, but so far it has been written in quite cumbersome coordinates. We would like to describe it by the metric \eqref{adschimetric}, but where $\phi$ has the periodicity $2\pi-\delta$, $\delta$ being the angular deficit. To go to this coordinate system, we will first apply the coordinate transformation discussed in Section \ref{geometry} to each wedge of allowed geometry (the regions bounded by $w_+^i$ and $w_-^{i+1}$). This brings us to a coordinate system consisting of $n$ static wedges, with parameters $(\nu_{i,i+1},\phi_{i,i+1},p_{i,i+1})$, with total deficit angle $\delta=2\pi-2\sum_i\nu_{i,i+1}$. 
This geometry is valid above the surface $\mathcal{L}$, and below $\mathcal{L}$ we still have empty AdS in the coordinates \eqref{adschimetric} (recall that this coordinate transformation maps $\mathcal{L}$ into itself). Now, by just pushing the wedges together and defining a continuous angular variable, the geometry above the shell becomes the metric \eqref{adschimetric} with periodicity $2\sum_i\nu_{i,i+1}$.\\

Let us call the coordinates below the shell $(\chi',t',\phi')$ and let the coordinates above the shell in the static conical singularity geometry be $(\chi,t,\phi)$. We would now like to know how these coordinates are related when we are on the light-like surface $\mathcal{L}$, in the limit $n\rightarrow\infty$. We thus define $\phi|_\mathcal{L}\equiv N(\phi')$ on $\mathcal{L}$, and it is easy to see that
\begin{equation}
N(\phi'_1)-N(\phi'_2)=\lim_{n\rightarrow\infty}\sum_{i,\psi_i\in (\phi'_1,\phi'_2)} 2\nu_{i,i+1}.
\end{equation}
Thus to fix the coordinate system we can choose $N(0)=0$, and we obtain $N(\phi')=\lim_{n\rightarrow\infty}\sum_{j,\psi_j<\phi'} 2\nu_{j,j+1}$. Let us also define continuous interpolating functions $Q(\psi_i)\equiv p_{i,i+1}\nu_{i,i+1}$ and $Z(\psi_i)\equiv\zeta_{i,i+1}$. In the limit we would like to express the three functions $Z$, $Q$ and $N$ in terms of $P$ and $\Phi$. From \eqref{zetadef} we can obtain $Z$ as
\begin{equation}
\tanh Z=\frac{P}{P\cos\Phi-\sin\Phi}+O(\frac{1}{n}).\label{Z}
\end{equation}
From \eqref{Gammaiis} and \eqref{Gammaii} we have in the limit $n\rightarrow\infty$ that
\begin{equation}
\tan Q\cosh Z=\frac{P-\tan\Phi}{1+P\tan\Phi}+O(\frac{1}{n}),\label{QZ}
\end{equation}
from which we can obtain $Q$. To obtain $N$, we first note that going to the subleading terms in equation \eqref{Gammaii} we can obtain
\begin{equation}
\tan\Gamma_+^{i,i+1}-\tan\Gamma_-^{i,i+1}=2\nu_{i,i+1}\frac{\cosh Z}{\cos^2Q}+O(\frac{1}{n^2}).
\end{equation}
Thus using this together with \eqref{tandiff}, we obtain
\begin{align}
N(\phi')=&\int_0^{\phi'}\frac{\cos^2Q(1+\tan^2\Phi)}{\cosh Z(1+P\tan\Phi)^2}\Big(-2\rho+P'+1+P^2\Big)\rd \phi.
\end{align}
We can actually simplify this expression a lot (see Appendix \ref{app3}), to obtain
\begin{equation}
N(\phi')=\int_0^\phi \frac{1-\cot\Phi P}{\cosh Z}.\label{anglerel}
\end{equation}
By using \eqref{Z} to obtain $Z$, this now gives us $N$ in terms of $P$ and $\Phi$, and thus indirectly in terms of $\rho$.\\

We are also interested in the relation between the radial coordinates when crossing the surface $\mathcal{L}$, namely we want to compute $\chi=\chi(\phi',\chi')$ on \LS. The mapping to the static coordinate system (before we push the circle segments together) is given by the equations \eqref{boosteqs}, and let's denote these coordinates by $(\tilde\phi,\chi,t)$ to distinguish them from $(\phi,\chi,t)$ which are the coordinates after we have pushed the circle segments together. Since light-like geodesics are mapped to light-like geodesics, we thus obtain from \eqref{boosteqs} that
\begin{equation}
\sinh\chi'=\sinh\chi(\cos Z+\sinh Z \cos(\phi_{i,i+1}-\tilde\phi)).\label{radeqpp}
\end{equation}
Note that, any point on \LS at an angle $\phi'\in(\psi_i,\psi_{i+1})$ will be mapped to a point with angle $\tilde\phi\in(\phi_{i,i+1}-(1-p_{i,i+1})\nu_{i,i+1},\phi_{i,i+1}+(1+p_{i,i+1})\nu_{i,i+1})$ (by definition of $\nu_{i,i+1}$ and $p_{i,i+1}$). In the limit, we thus have that $\psi_i$ is mapped to $\phi_{i,i+1}+Q(\psi_i)$, thus we have from equation \eqref{radeqpp} that the radial coordinates are related on $\mathcal{L}$ as
\begin{equation}
\sinh\chi=\frac{\sinh\chi'}{\cosh Z(\phi')+\sinh Z(\phi')\cos Q(\phi')}=\frac{\sinh\chi'}{N'(\phi')}.\label{radeqpp2}
\end{equation}
See Appendix \ref{app3} for the second equality. The fact that this proportionality factor is the inverse of the function relating the angles in \eqref{anglerel},  turns out to be a requirement for a well defined thin shell spacetime.\\

We would now like to bring the geometry to an \ads-Vaidya type of geometry. Let us first define $A=\frac{1}{\pi}\sum_i\nu_{i,i+1}=\frac{1}{2\pi}N(2\pi)$, which thus is the total angle of the conical singularity divided by $2\pi$. We now want to find coordinate transformations such that 
\begin{equation}
\rd s_\pm^2=-f_\pm\rd v_\pm^2+2\rd v_\pm\rd r_\pm+r_\pm^2\rd \phi_\pm^2,\label{vaidya}
\end{equation}
where $+$ is is the patch after the shell and $-$ is before the shell, and the light-like boundary $\mathcal{L}$ between the two patches is given by  $v_\pm=0$. We have $f_-=1+r_-^2$ (empty \ads) and $f_+=A^2+r_+^2$, and we want the periodicity of $\phi_\pm$ to be $2\pi$. To do this, we first do the coordinate transformation $\sinh \chi'=r_-$, $t'=t_-$ and $\phi'=\phi_-$ in the empty \ads part, and for the conical singularity part we have $\sinh\chi=r_+/A$, $t=A t_+$ and $\phi=A\phi_+$. This gives us the metric
\begin{equation}
\rd s_\pm^2=-f_\pm\rd t_\pm^2+\rd r_\pm^2/f_\pm+r_\pm^2\rd \phi_\pm^2.
\end{equation}
From this we obtain the metric \eqref{vaidya} by the standard coordinate transformation to infalling coordinates, given by $\rd v_\pm=\rd t_\pm+\rd r_\pm/f_\pm$. The relations between the radial and angular coordinates when crossing the shell can now be written as
\begin{equation}
r_+=G(\phi_-)r_-\label{rpm}
\end{equation}
\begin{equation}
\phi_+=\int_0^{\phi_-}\frac{1}{G(\phi)}\rd\phi.
\end{equation}
where
\begin{equation}
G(\phi_-)=\frac{A}{\cosh Z(\phi_-)+\sinh Z(\phi_-)\cos Q(\phi_-)}=\frac{A\cosh Z(\phi_-)}{1-P(\phi_-)\cot\Phi(\phi_-)}.
\end{equation}
The fact that the proportionaly factor relating the radial coordinates turns out to be exactly the inverse of the derivative of the function relating the angular coordinates, is a necessary (and sufficient) condition to have a well defined induced metric on the shell, which is important when analyzing the junction conditions in Section \ref{thinshell}, and can be seen as a non-trivial consistency check. It should be mentioned that these thin shell solutions are probably unphysical, since the thin shell limit of a thick shell composed of some dynamical matter (for example a scalar field) is expected to scatter below the black hole threshhold, and not form a conical singularity. Nevertheless, these solutions can be constructed formally.\\

It is instructive to look at the rotationally symmetric case, where $\rho=\mathrm{const.}$, which should reduce to the standard AdS-Vaidya spacetime for a conical singularity. Thus we let $e_i=e=\tan\epsilon$, and we have $\Gamma^i_\pm=\Gamma_\pm=\pm\epsilon$, as well as $\nu_{i,i+1}=\nu=A\rd\phi/2$. Defining $e=\rho \rd\phi$ we obtain from equation \eqref{zetadef} that
\begin{equation}
\tanh\zeta=\frac{2\rho}{2\rho-1},
\end{equation}
and from \eqref{Gammaii} and \eqref{Gammaiis} we obtain
\begin{equation}
1-2\rho=A\cosh\zeta.
\end{equation}
The condition that $-1\leq\tanh\zeta\leq1$ and $\rho\geq0$ gives us that $\rho\leq1/4$ which thus is the threshhold for black hole formation. We can solve for $A$ in terms of $\rho$ to obtain the simple relation
\begin{equation}
A=\sqrt{1-4\rho}.
\end{equation}
At the threshhold of black hole formation we have $A\rightarrow0$, meaning that the total angle of the conical singularity approaches zero. The relation between $r_+$ and $r_-$ can be obtained from equation \eqref{rpm} and we obtain that in this case $r_+=r_-$ when crossing the shell.

\subsection{Black hole}
In the black hole case the situation is a little bit less intuitive than in the formation of a massive particle. In this case, we would like to transform the metric after the shell into the form \eqref{BTZ}. We proceed as in the point particle case and first transform the wedges of allowed geometry to wedges of the form \eqref{btzeq0}. Then we want to go to the coordinate system described in Section \ref{secSch}, where we in particular let $x_0=\rho \cosh y$ and $x_2=\rho\sinh y$, which will take each wedge to be given by a circle segment of \eqref{BTZ}. The total black hole will then be trivially constructed by gluing together all these static wedges, as in Section \ref{secMerge}.\\.

Recall that wedges of the form \eqref{btzeq0} with parameters $\mu_\pm=\mu\pm\xi$ will be mapped to wedges at constant $y=\xi\pm\mu$ in the coordinate system with metric \eqref{BTZ}. The coordinates below the shell will be denoted by $(\chi',t',\phi')$, and the coordinates after the shell with metric \eqref{BTZ} will be $(r,t,y)$. Analogously with the case of a pointlike particle, we assign the wedge of allowed geometry bounded by $w_+^i$ and $w_-^{i+1}$ the parameters $\mu_{i,i+1}^\pm=\mu_{i,i+1}\pm\xi_{i,i+1}$, and it thus follows that points on the light-like surface $\mathcal{L}$ at an angle $\phi'$ in the interval $(\psi_i,\psi_{i+1})$ will be mapped to points with a $y$ value in the inteval $(\xi-\mu,\xi+\mu)$ (the circle segment corresponding to this interval thus has angle $2\mu$). Analogously with the formation of a pointlike particle, we thus define $y(\phi')|_{\mathcal{L}}\equiv N(\phi')$, and we see that
\begin{equation}
N(\phi'_1)-N(\phi'_2)=\lim_{n\rightarrow\infty}\sum_{i,\psi_i\in (\phi'_1,\phi'_2)} 2\mu_{i,i+1}.
\end{equation}
We can again fix the coordinate system such that $y(0)|_{\mathcal{L}}=N(0)=0$, so that $N(\phi')=\lim_{n\rightarrow\infty}\sum_{j,\psi_j<\phi'} 2\mu_{j,j+1}$. Let us also define continuous interpolating functions $X(\psi_i)\equiv \xi_{i,i+1}$ and $Z(\psi_i)\equiv\zeta_{i,i+1}$. In the limit we would like to express the three functions $Z$, $X$ and $N$ in terms of $P$ and $\Phi$. From \eqref{zetadef} we can obtain $Z$ as
\begin{equation}
\coth Z=\frac{P}{P\cos\Phi-\sin\Phi}+O(\frac{1}{n}).\label{Zbh}
\end{equation}
From \eqref{Gammaiibh} and \eqref{Gammaiis} we have in the limit $n\rightarrow\infty$ that
\begin{equation}
-\tanh X\sinh Z=\frac{P-\tan\Phi}{1+P\tan\Phi}+O(\frac{1}{n}),\label{QX}
\end{equation}
from which we can obtain $X$. Now from equation \eqref{Gammaiibh} we obtain
\begin{equation}
\tan \Gamma_+^{i,i+1}-\tan \Gamma_-^{i,i+1}=-\frac{2}{\cosh^2\xi_{i,i+1}}\mu_{i,i+1}\sinh\zeta_{i,i+1},
\end{equation}
and just like in the massive particle case, we can use this result together with \eqref{tandiff}, to obtain
\begin{equation}
N(\phi')=-\int_0^{\phi'}\frac{\cosh^2X(1+\tan^2\Phi)}{\sinh Z(1+P\tan\Phi)^2}\Big(-2\rho+P'+1+P^2\Big)\rd \phi.
\end{equation}
This can be simplified to (see Appendix \ref{app3})\\
\begin{equation}
N(\phi')=\int_0^\phi \frac{P\cot\Phi-1}{\sinh Z}.\label{anglerelbh}
\end{equation}

 Now we want to figure out how the radial coordinate before the shell $\chi'$ is related to the radial coordinate $\rho$, so consider the wedge bounded by $w_+^i$ and $w_-^{i+1}$. Let $(\chi,t,\phi)$ be the coordinate system after the shell in which a wedge takes the form \eqref{btzeq0}. From \eqref{btzeq0} we obtain that light-like geodesics with equation $\tanh\chi'=-\sin t'$ at angle $\psi_i$ and $\psi_{i+1}$ are mapped to light-like geodesics at an angle $\phi$ given by $\sin\phi=\pm\tanh (\mu\pm\xi)$ with the $+$ ($-$) for $\psi_{i+1}$ ($\psi_i$). In the limit, we thus obtain that all light-like geodesics at an angle in $(\psi_i,\psi_{i+1})$ are mapped to a light-like geodesic at an angle $\phi$ given by $\sin\phi=\tanh\xi_{i,i+1}=\tanh X$. Now, from \eqref{boosteqs} we obtain that the radial coordinates $\chi$ and $\chi'$ are related by
\begin{equation}
\sinh\chi=\frac{\sinh\chi'}{\cosh Z+\sinh Z\cos \phi}=\frac{\sinh\chi'}{\cosh Z+\frac{\sinh Z}{\cosh X}}.
\end{equation}
Now, it is easy to see that the light-like geodesics satisfy $x_1^2+x_2^2=x_0^2$, and we can obtain $\rho$ by
\begin{equation}
\rho^2=x_0^2-x_2^2=\sinh^2\chi\cos\phi^2=\frac{\sinh^2\chi}{\cosh^2X}\Rightarrow \rho=\frac{\sinh\chi'}{\cosh X\cosh Z+\sinh Z}=\frac{\sinh\chi'}{N'(\phi')},\label{radeqbh}
\end{equation}
See Appendix \ref{app3} for the last equality. To make the angular coordinate $y$ have period $2\pi$, we may rescale $y\rightarrow y\sqrt{M}$, $\rho\rightarrow \rho/\sqrt{M}$ and $t\rightarrow t\sqrt{M}$, where the mass is given by $M=\frac{1}{\pi^2}\left(\sum_i\mu_{i,i+1}\right)^2$, which will give us exactly the metric \eqref{BTZ}. To summarize, our resulting spacetime can be written on the form
\begin{equation}
\rd s_\pm^2=-f_\pm\rd v_\pm^2+2\rd v_\pm\rd r_\pm+r_\pm^2\rd \phi_\pm^2\label{vaidya2}
\end{equation}
where + (-) is after (before) the shell, while $f_+=-M+r_+^2$ and $f_-=1+r_-^2$. The coordinates are related on the shell as
\begin{equation}
r_+=r_-G(\phi_-),
\end{equation}
\begin{equation}
\phi_+=\int_0^{\phi_-}\frac{1}{G(\phi)}\rd\phi,
\end{equation}
where
\begin{equation}
G(\phi)=\frac{\sqrt{M}}{\cosh X\cosh Z+\sinh Z}=\frac{\sqrt{M}\sinh Z}{P\cot\Phi-1}.\label{Gdefbh}
\end{equation}

Let us again consider the case of rotational symmetry. We thus define $e_i=e=\rho\rd\phi$ where $\rho=$ is constant, and again we have $\Gamma_\pm^i=\pm\epsilon$. We also have $\mu_{i,i+1}=\mu=\sqrt{M}\rd\phi/2$, and we obtain
\begin{equation}
 \coth Z=\frac{2\rho}{2\rho-1},
\end{equation}
and
\begin{equation}
 2\rho-1=\sqrt{M}\sinh Z.
\end{equation}
We can solve for $M$ as
\begin{equation}
M=4\rho-1.
\end{equation}
We also obtain that $r_+=r_-$ when crossing the shell.

\section{Thin shell formalism}\label{thinshell}
Here we will explore the thin shell formalism for light-like shells, and we will use this to determine the stress-energy tensor of the shell obtained from the point particle approach. We will thus consider two spacetimes $\R_-$ and $\R_+$, with metrics
\begin{equation}
\rd s_\pm^2=-f_\pm\rd v_\pm^2+2\rd r_\pm\rd v_\pm+r_\pm^2\rd \phi_\pm^2,
\end{equation}
which are glued together along a light-like surface. We let $f_-=1+r_-^2$ and $f_+=\Omega+r_+^2$ ($\Omega=A^2$ for the creation of a conical singularity and $\Omega=-M$ for the creation of a black hole). The light-like surface $\mathcal{L}$ which will separate the two regions is given by $v_\pm=0$. The normal vectors of the shell, which satisfy $n\cdot n=0$ and are orthogonal to the tangent space of the shell are given by
\begin{equation}
n_\pm^\mu=a_\pm(0,-1,0)
\end{equation}
for some $a_\pm$. These are determined solely in terms of the embedding of the null surface, and does not say anything about the matter content of the shell (unlike the case with massive shells\cite{Israel:1966rt}\cite{Musgrave:1995ka}). However, there will still be a non-trivial junction, which is determined completely in terms of the relation between the coordinates across the shell, which reads $r_+=G(\phi_-)r_-$ for some function $G(\phi)$, and as we will see, to be able to define consistently a metric on the shell we must have $\phi_+=H(\phi_-)$ where $H'(\phi_-)=1/G(\phi_-)$ (which is exactly what we obtained from the point particle construction and is a good consistency check). We will fix $a_+$ in terms of $a_-$ by requiring that $n$, when projected onto the shell, is the same from each side of the shell. We could then fix $a_-$ to be say equal to one, but we will keep it through the whole calculation and we will explicitly see in the end that the final result is independent of $a_-$.\\

The shell will be parametrized by two coordinates, and for convenience we will choose the coordinates $\phi\equiv\phi_-$ and $r\equiv r_-$, such that the embedding in $\R_-$ is simple (but it will be more complicated in $\R_+$). The basis vectors of the tangent space on the shell are in $\R_-$ given by
\begin{equation}
e_{r}^-=\frac{\partial x_-^\alpha}{\partial r}=(0,1,0),
\end{equation}
\begin{equation}
e_{\phi}^-=\frac{\partial x_-^\alpha}{\partial \phi}=(0,0,1),
\end{equation}
and in $\R_+$ we have
\begin{equation}
e_{r}^+=\frac{\partial x_+^\alpha}{\partial r}=(0,G(\phi),0),
\end{equation}
\begin{equation}
e_{\phi}^+=\frac{\partial x_+^\alpha}{\partial \phi}=(0,G'(\phi)r,1/G(\phi)).
\end{equation}
The (degenerate) metric $g_{ab}=e^{-\alpha}_{a}e^{-\beta}_{b}g_{\alpha\beta-}=e^{+\alpha}_{a}e^{+\beta}_{b}g_{\alpha\beta+}$ on the shell is then
\begin{equation}
\rd s^2=r^2\rd \phi^2.
\end{equation}
As mentioned before, the metric is well defined (or in other words, we obtain the same induced metric from both sides) only if $\frac{d\phi_+}{d\phi_-}=\frac{r_-}{r+}$. Note that since the normal vector $n$ is proportional to $e_r$, we will now fix $a_+=a_- G(\phi)=$ such that the proportionality factor is the same and $n$ is the ``same'' vector on each side of the shell.  \\

To determine the shells stress-energy tensor, we will follow the procedure in \cite{Barrabes:1991ng}\cite{Musgrave:1997}. Since the normal vectors for nullike shells do not give any information about the gluing, the null shell formalism requires that we define two transverse vectors $N_\pm$ given by the conditions $[N_\alpha e^\alpha_a]=0$ and $[N\cdot N]=0$, and such that $N\cdot n\neq 0$. For convenience we choose $N\cdot n=-1$, and this completely determines $N$ up to shifts of tangential vectors. We will thus pick
\begin{equation}
N_-^\mu=\frac{1}{a_-}(1,0,0),
\end{equation}
where we have used the freedom of tangential shifts to set the $\phi$ and $r$ components of $N_-$ to zero (by convention $N$ points away from $\R_-$ and into $\R_+$, which would require $a_->0$). It then follows that
\begin{equation}
N_+^\mu=(\frac{1}{a_-G},\frac{1}{2a_-G}f_+-\frac{G}{2a_-}f_--\frac{(G')^2}{2a_-G},-\frac{G'}{G^2ra_-}).
\end{equation}
Due to the degeneracy of the metric, the junction formalism relies on the so called generalized extrinsic curvatures, which follow the same definition as the standard extrinsic curvatures but with the normal vector replaced by an arbitrary vector. Thus the generalized extrinsic curvatures corresponding to the vectors $N$ are defined as 
\begin{equation}
K_{ab}\equiv-N_\mu e^\nu_{(a)}\nabla_\nu e^\mu_{(b)}=-N_\mu\left(\frac{\partial^2x^\mu}{\partial \xi^{(a)}\partial \xi^{(b)}}+\Gamma^\mu_{\alpha\beta}e^\alpha_{(a)}e^\beta_{(b)}\right).
\end{equation}
To compute these we need the Christoffel symbols, which are
\begin{align}
\begin{array}{lll}
 \Gamma^v_{vv}=\frac{f'_\pm}{2},&\Gamma^v_{\phi\phi}=-r_\pm,&\Gamma^r_{vv}=\frac{f_\pm f'_\pm}{2},\\
 \Gamma^r_{vr}=-\frac{f_\pm'}{2},&\Gamma^r_{\phi\phi}=-r_\pm f_\pm,&\Gamma^\phi_{r\phi}=\frac{1}{r_\pm},
\end{array}
\end{align}
where all other components vanish (we dropped some of the $\pm$ subscripts on the indices in the above formulas). The only non-zero extrinsic curvature component turns out to be
\begin{align}
K^+_{\phi\phi}=&\frac{rf_+}{2a_-G^2}-\frac{rf_-}{2a_-}-\frac{rG''}{a_-G}+\frac{r(G')^2}{2a_-G^2}\nonumber\\
=&\frac{r}{2a_-}\left(\frac{\Omega}{G^2}-1\right)-\frac{rG''}{a_-G}+\frac{r(G')^2}{2a_-G^2}.\label{Keq}
\end{align}
We now define $\gamma_{ij}=2[K_{ij}]$, which thus has the only non-zero component $\gamma_{\phi\phi}=2K^+_{\phi\phi}$.

\subsection{Intrinsic formulation}
Now the intrinsic stress-energy tensor of the shell is uniquely determined by $\gamma_{ij}$. However, to get there, we will first define a method for raising indices on the shell. Since the metric is degenerate, we can not use it for this purpose, and instead indices are raised by a different tensor. To construct this tensor, we first decompose the normal vector in terms of the basis $(N,e)$ as
\begin{equation}
n^\alpha=\ell^a e_a^\alpha,
\end{equation}
from which we immediately see that
\begin{equation}
\ell^r=-a,\textrm{   } \ell^\phi=0.
\end{equation}
Indices are now raised by the quantity $g_*^{ab}$ defined by
\begin{equation}
g_*^{ac}g_{cb}=\delta_b^a-\eta \ell^a e^\alpha_{(b)} N_\alpha=\left(\begin{array}{cc} 0 & 0 \\ 0 & 1 \\ \end{array}\right).
\end{equation}
$g_*$ is not uniquely defined, but we can choose
\begin{equation}
g^{ab}_*=\left(\begin{array}{cc} 0 & 0 \\ 0 & \frac{1}{r^2} \\ \end{array}\right).
\end{equation}
The instrinsic stress-energy tensor of the shell is now given by
\begin{equation}
-16\pi S^{ab}=\left(g_*^{ac}\ell^b\ell^d+\ell^a\ell^cg_*^{bd}-g_*^{ab}\ell^c\ell^d-\ell^a\ell^bg_*^{cd}\right)\gamma_{cd}.
\end{equation}
From which we obtain the only non-zero component of the stress-energy tensor as
\begin{equation}
16\pi S^{rr}=\frac{a_-^2}{r^2}\gamma_{\phi\phi}.
\end{equation}
Note that the result depends on the arbitrary normalization factor $a_-$ (which is not even assumed to be constant). To obtain a result independent of this factor, we must compute the extrinsic stress-energy tensor as it appears in the Einstein equations.
\subsection{Extrinsic formulation}
Th extrinsic stress-energy tensor of the shell, $S^{\mu\nu}$, is related to the intrinsic one by $S^{\mu\nu}=S^{ij}e_i^\mu e_j^\nu$, but for instructive reasons we will here compute it using the algorithm outlined in \cite{Barrabes:1991ng}\cite{Musgrave:1997}. We will for concreteness only look at how the stress-energy tensor looks like in the coordinates below the shell. We first need to define a tensor $\gamma_{\mu\nu}$, whose projection onto the surface $\mathcal{L}$ is $\gamma_{ij}$. It is easy to see that $\gamma_{\mu\nu}$ is given by $\gamma_{\mu\nu}|_{\mu=\phi,\nu=\phi}=\gamma_{ij}|_{i=\phi,j=\phi}$ and all other components vanish. The extrinsic stress-energy tensor is then given by
\begin{equation}
-16 \pi S^{\mu\nu}=2\gamma^{(\mu}n^{\nu)}-\gamma n^\mu n^\nu-\gamma^\dagger g^{\mu\nu},
\end{equation}
where $\gamma^\alpha=\gamma^{\alpha\beta}n_\beta$, $\tilde{\gamma}=\gamma^\alpha n_\alpha$ and $\gamma=\gamma_{\alpha\beta}g^{\alpha\beta}$. We see that $\gamma^\alpha=0$ and $\tilde{\gamma}=0$ while $\gamma=\gamma_{\phi\phi}r^{-2}$. Thus we obtain that the only non-zero component of $S^{\alpha\beta}$ is
\begin{equation}
16\pi S^{rr}=\frac{a_-^2}{r^2}\gamma_{\phi\phi}.
\end{equation}
and we see indeed that we have $S^{\mu\nu}=S^{ij}e_i^\mu e_j^\nu$.\\

The stress-energy tensor in the enveloping spacetime is now given by
\begin{equation}
T^{\mu\nu}=\alpha S^{rr}\delta(F),
\end{equation}
where $F=0$ defines the shell and the normal vector is given by $(n_-)_\mu=\frac{1}{\alpha}\partial_\mu F$ (which defines $\alpha$). Choosing $F=v_-$ (and thus $\alpha=-1/a_-$) we obtain the only non-zero component as
\begin{equation}
T^{rr}=-\frac{a_-}{16\pi r^2}\gamma_{\phi\phi}\delta(v_-).
\end{equation}
Note that since $\gamma_{\phi\phi}\sim1/a_-$, this result will be independent of the arbitrary normalization $a_-$ as expected. The dependence on the normalization $a_-$ in the stress-energy tensor of the shell can thus be understood as an ambiguity in the delta function for the stress-energy tensor as it appears in Einstein's equations.
\subsection{Connection with colliding pointlike particles}
So far we have defined our thin shell spacetime by the function $G(\phi_-)$. We will now assume that this function is the result of a limit of solutions of colliding pointlike particles. In this case (see Appendix \ref{app4}) we obtain the simple result
\begin{equation}
\frac{a_-}{r}K^+_{\phi\phi}=-2\rho(\phi_-).\label{awesome}
\end{equation}
Recall that $\rho$ is the continuous distribution of the pointlike particles, namely such that $e_i=\rho(\psi_i)\rd\phi$ for large finite $n$ in the limit. The stress-energy tensor is then given by (recall $\gamma_{\phi\phi}=2K_{\phi\phi}$)
\begin{equation}
T^{rr}=-\frac{a_-}{16\pi r^2}\gamma_{\phi\phi}\delta(v_-)=\frac{\rho(\phi_-)}{4\pi r}\delta(v_-).
\end{equation}
Thus the energy density computed by the thin shell formalism is proportional to the distribution of the pointlike particles, which is expected but is still a very non-trivial consistency check.

\section{Conclusions and outlook}\label{secConclusions}
We have studied the formation of a black hole or a conical singularity from an arbitrary number of pointlike particles. We have showed how to construct a consistent coordinate system for this spacetime, which is a non-trivial task when going away from discrete rotational symmetry. We also showed that it is possible to take the limit of an infinite number of particles to obtain a thin shell spacetime, and we computed the energy density using the thin shell formalism and obtained agreement with the expectation from the pointlike particle picture. One of the most interesting results is that we have obtained thin shell spacetimes that break rotational symmetry, which seems to not have been discussed in the literature before. These thin shell spacetimes can be defined without a reference to pointlike particles, and are a generalization of the AdS$_3$-Vaidya spacetimes, but where there is a non-trivial angle-dependent mapping between the radial and angular coordinates when crossing the shell. According to the AdS/CFT correspondence, these solutions correspond to an instantaneous perturbation (injection of energy) in the dual CFT that has some angular dependence, and studying field theory observables (such as entanglement entropies) in these spacetimes will be an interesting topic for future research (the homogeneous case has been studied for the Poincar\'e patch in \cite{Balasubramanian:2011ur}, and recently for global coordinates in \cite{Ziogas:2015aja}). In higher dimensions one would not expect to find such simple examples of spacetimes that break rotational symmetry since in higher dimensions we would expect that dynamical gravitational modes will be excited. \\

An interesting extension of the construction in this paper would be to add angular momentum. For this it might be possible to collide spinning pointlike particles, and the resulting object is then expected to be a spinning massive particle or a rotating BTZ black hole. In the thin shell limit, we expect to have another continuous function specifying the angular momentum density of the boundary source. Relaxing the assumption that the particles are created at equal times is also an interesting problem, although this makes the setup a lot more complicated (it would also be necessary to have non-radial geodesics for the particles). In the thin shell limit this would result in solutions corresponding to a source that is turned on at different times at the boundary. We leave these questions to future research. Other possible research directions would be to study the formation of charged BTZ black holes from charged pointlike particles (or maybe even higher spin black holes), but it is not clear how this would work since the charged BTZ black hole can not be obtained by just identifying points in \ads, and a more clever construction is thus needed. Understanding the construction in this paper using the Chern-Simons formulation of three-dimensional gravity might be helpful for understanding such constructions.
\section{Acknowledgements}
The author would like to thank Ben Craps for comments on the manuscript. This work was supported in part by the Belgian Federal Science Policy Office through the Interuniversity Attraction Pole P7/37, by FWO-Vlaanderen through project G020714N, and by the Vrije Universiteit Brussel through the Strategic Research Program ``High-Energy Physics''. The author is supported by a PhD fellowship from the Research Foundation Flanders (FWO); his work was also partially supported by the ERC Advanced Grant ``SyDuGraM", by IISN-Belgium (convention 4.4514.08) and by the ``Communaut\'e Fran\c{c}aise de Belgique" through the ARC program.

\appendix
\section{Derivation of equation \eqref{c1}}\label{app1}

The last three equations in \eqref{boosteqs} can be written as
\begin{equation}
\sin t'\cosh \chi'=\cosh\chi\sin t \cosh \zeta + \sinh\chi\sinh\zeta\sin(\alpha-\phi)\label{appeq1},
\end{equation}
\begin{equation}
\sinh\chi'\cos\phi'=\cosh\chi\sin t\sinh\zeta\sin\alpha+\sinh\chi(\cos\alpha\cos(\phi-\alpha)-\cosh\zeta\sin\alpha(\phi-\alpha)\sin\alpha)\label{appeq2},
\end{equation}
\begin{equation}
\sinh\chi'\sin\phi'=-\cosh\chi\sin t\sinh\zeta\cos\alpha+\sinh\chi(\sin\alpha\cos(\phi-\alpha)+\cosh\zeta\sin(\phi-\alpha)\cos\alpha)\label{appeq3}.
\end{equation}
We would like to take linear combinations of the two last equations and make them proportional to the first one, thus we write $\sin A$\eqref{appeq2}+$\cos A$\eqref{appeq3}$=K$\eqref{appeq1}. Evaluating this we obtain

\begin{align}
\sinh\chi'\sin(\phi'+A)=&-\cosh\chi\sin t\sinh\zeta\cos(\alpha+A)+\sinh\chi\sin(\alpha+A)\cos(\phi-\alpha)+\nonumber\\
 &+\sinh\chi\cosh\zeta\cos(\alpha+A)\sin(\phi-\alpha)\nonumber\\
 =&K\left(\cosh\chi\sin t\cosh\zeta+\sinh\chi\sinh\zeta\sin(\alpha-\phi)\right)\nonumber\\
  =&K\sin t' \cosh \chi'\label{appeq4}
\end{align}
By comparing the coefficients of $\cosh\chi\sin t$ and $\sinh\chi$ we thus obtain the equation
\begin{align}
\sin(\alpha-\phi)\tanh\zeta&=-\sinh^{-1}\zeta\tan(\alpha+A)\cos(\phi-\alpha)-\coth\zeta\sin(\phi-\alpha)\nonumber\\
&\Rightarrow \tan(\alpha+A)=\frac{\tan(\alpha-\phi)}{\cosh\zeta},
\end{align}
and $K$ is then given as
\begin{equation}
K=-\tanh\zeta\cos(\alpha+A).
\end{equation}
Now we let $\phi=\phi_\pm=\psi\pm(1\pm p)\nu$, $\alpha=\psi-\pi/2$ and $\Gamma_\pm\equiv -\psi-A$, which gives
\begin{equation}
K=\tanh\zeta\sin\Gamma_\pm,
\end{equation}
and
\begin{equation}
\tan\Gamma_\pm=-\cosh\zeta\tan(\psi-\phi_\pm)=\pm\cosh\zeta\tan((1\pm p)\nu),
\end{equation}
and equation \eqref{appeq4} reduces exactly to \eqref{c1}.

\section{Derivation of equation \eqref{peq}}\label{app2}
We parametrize the intersection $I_{i-1,i}$ between $w_-^i$ and $w_+^{i-1}$ as
\begin{equation}
I_{i-1,i}=\cosh\chi'\sin t'\gamma_0+\sinh\chi'\cos\phi_{i-1,i}\gamma_1+\sinh\chi'\sin\phi_{i-1,i}\gamma_2+\cosh\chi'\cos t',
\end{equation}
and the intersection $I_{i,i+1}$ between $w_+^i$ and $w_-^{i+1}$ as
\begin{equation}
I_{i,i+1}=\cosh\chi\sin t\gamma_0+\sinh\chi\cos\phi_{i,i+1}\gamma_1+\sinh\chi\sin\phi_{i,i+1}\gamma_2+\cosh\chi\cos t.
\end{equation}
Computing $u_i^{-1}I_{i,i+1}u_i=I_{i-1,i}$ gives us the equations
\begin{subequations}
\begin{align}
\sin t'\cosh\chi'=&2e_i^2\sin t\cosh\chi+2e_i^2\cos(\phi_{i,i+1}-\psi_i)\sinh\chi\nonumber\\
&-2e_i\sin(\phi_{i,i+1}-\psi_i)\sinh\chi+\sin t\cosh\chi,\label{weqsa}\\
\cosh\phi_{i-1,i}\sinh\chi'=&-2e_i^2\sin\psi_i\sin(\phi_{i,i+1}-\psi_i)\sinh\chi-2e_i^2\sin t\cos\psi_i\cosh\chi\nonumber\\
&-2e_i^2\cos\phi_{i,i+1}\sinh\chi+2e_i\sin\phi_{i,i+1}\sinh\chi\nonumber\\
&+2e_i\sin\psi_i\sin t\cosh \chi+\cos\phi_{i,i+1}\sinh\chi\label{weqsb},\\
\sin\phi_{i-1,i}\sinh\chi'=&-2e_i^2\sin\psi_i\sin t\cosh\chi-2e_i^2\sin\psi_i\cos(\phi_{i,i+1}-\psi_i)\sinh\chi\nonumber\\
&-2e_i\sin t\cos\psi_i\cosh\chi-2e_i\cos\phi_{i,i+1}\sinh\chi+\sin\phi_{i,i+1}\sinh\chi,\label{weqsc}\\
\cos t\cosh\chi=&\cos t'\cosh\chi'.\label{weqsd}
\end{align}
\end{subequations}
The above equations correspond to equating the coefficients of $\gamma_0$, $\gamma_1$, $\gamma_2$ and $\mathbf{1}$, respectively. We also have the following equations that must be satisfied
\begin{subequations}
\begin{align}
\tanh\chi'\sin(-\phi_{i-1,i}+\Gamma_-^i+\psi_i)=&-\sin\Gamma_-^i\sin t'\label{weqconsta},\\
\tanh\chi\sin(-\phi_{i,i+1}+\Gamma_+^i+\psi_i)=&-\sin\Gamma_+^i\sin t\label{weqconstb},
\end{align}
\end{subequations}
where
\begin{equation}
\tan\Gamma_{\pm}^i=\pm(1\pm p_i)e_i.
\end{equation}
The goal is now to find $p_i$ such that all these equations are satisfied. First, we can eliminate $t'$ and $t$ in terms of $\chi'$ and $\chi$, respectively, which only leaves us the coordinates $\chi$ and $\chi'$. Then we must specify $p_i$, such that, for every $\chi$\footnote{That is, $\chi$ is a free variable, but there might be certain bounds for $\chi$.}, we can find a $\chi'$ such that \eqref{weqsa}-\eqref{weqsd} are satisfied. Now we use \eqref{weqconsta} and \eqref{weqconstb} to eliminate $\cosh\chi\sin t$ in \eqref{weqsc}, to obtain
\begin{equation}
\sinh\chi'=\frac{-(e_ip_i\cos\phi_{i,i+1}-e_ip_i\cos(\phi_{i,i+1}-2\psi_i)-p_i\sin\phi_{i,i+1}+\sin(\phi_{i,i+1}-2\psi_i))\sinh\chi}{(p_i+1)\sin\phi_{i-1,i}}.
\end{equation}
Now it is convenient to substitute this into the linear combination \eqref{weqsb}$-e_i$\eqref{weqsc}, from which we can then solve for $p_i$ as
\begin{equation}
p_i=\frac{\sin(\phi_{i,i+1}+\phi_{i-1,i}-2\psi_i)}{-e_i\cos(\phi_{i,i+1}-\phi_{i-1,i})+e_i\cos(\phi_{i,i+1}+\phi_{i-1,i}-2\psi_i)+\sin(\phi_{i,i+1}-\phi_{i-1,i})},
\end{equation}
which is equivalent to \eqref{peq}. To obtain this result we only used equations \eqref{weqsb} and \eqref{weqsc}. It is now possible to explicitly check that equation \eqref{weqsa} is satisfied, although the computations are quite tedious. We then know that equation \eqref{weqsd} must be satisfied, because the equations are not independent (or to be more precise, equations \eqref{weqsa}-\eqref{weqsc} imply via the embedding equation (or condition of unit determinant) that $\cos t\cosh\chi=\pm\cos t'\cosh\chi'$, and the condition to have the plus sign just tells us in what range we have to pick $t'$). 

%

\section{Derivation of equations \eqref{anglerel}, \eqref{radeqpp2}, \eqref{anglerelbh} and \eqref{radeqbh}}\label{app3}
Starting with
\begin{equation}
N'(\phi')=\frac{\cos^2Q(1+\tan^2\Phi)}{\cosh Z(1+P\tan\Phi)^2}\Big(-2\rho+P'+1+P^2\Big),
\end{equation}
we first use equation \eqref{ode1} to write this as
\begin{align}
N'(\phi')=\frac{\cos^2Q(1+\tan^2\Phi)}{\cosh Z(1+P\tan\Phi)^2}(1-\cot\Phi P),
\end{align}
From equation \eqref{Z} we have
\begin{equation}
\frac{1}{\cosh^2 Z}=1-\tanh^2Z=1-\frac{P^2}{(P\cos\Phi-\sin\Phi)^2},
\end{equation}
and we then obtain from \eqref{QZ} that
\begin{align}
\frac{1}{\cos^2Q}=1+\tan^2Q=&1+\frac{(P-\tan\Phi)^2}{(1+P\tan\Phi)^2}\frac{1}{\cosh^2Z}\nonumber\\
=&1+\frac{-P^2+(P\cos\Phi-\sin\Phi)^2}{(\cos\Phi+P\sin\Phi)^2}\nonumber\\
=&\frac{1}{(\cos\Phi+P\sin\Phi)^2}.
\end{align}
From this we immediately obtain
\begin{align}
N'(\phi')=\frac{1-\cot\Phi P}{\cosh Z}.
\end{align}
We would now like to derive the second equality in equation \eqref{radeqpp2}. Since $\tanh Z=\frac{P}{P\cos\Phi-\sin\Phi}$, we have $\sinh Z\cosh Z=\frac{P(P\cos\Phi-\sin\Phi)}{-P^2+(P\cos\Phi-\sin\Phi)^2}$ and $\cosh^2 Z=\frac{(P\cos\Phi-\sin\Phi)^2}{-P^2+(P\cos\Phi-\sin\Phi)^2}$, and we obtain that
\begin{align}
\frac{1}{\cosh Z+\sinh Z\cos Q}=&\cosh Z\frac{-P^2+(P\cos\Phi-\sin\Phi)^2}{(P\cos\Phi-\sin\Phi)^2+P(P\cos\Phi-\sin\Phi)(\cos\Phi+P\sin\Phi)}\nonumber\\
=&\frac{\cosh Z}{1-P\cot\Phi}.
\end{align}
which proves \eqref{radeqpp2}, and where we also used $\cos Q=\cos\Phi+P\sin\Phi$ since it can be seen that $\cos\Phi+P\sin\Phi>0$.\\

For the black hole case, the situation is very similar. We then start with
\begin{equation}
N'(\phi')=-\frac{\cosh^2X(1+\tan^2\Phi)}{\sinh Z(1+P\tan\Phi)^2}(1-\cot\Phi P),\label{Npbh}
\end{equation}
where we again used \eqref{ode1}. From equation \eqref{Zbh} we have
\begin{equation}
\frac{1}{\sinh^2 Z}=\coth^2Z-1=\frac{P^2}{(P\cos\Phi-\sin\Phi)^2}-1,
\end{equation}
and we then obtain from \eqref{QX} that
\begin{align}
\frac{1}{\cosh^2X}=1-\tanh^2X=&1-\frac{(P-\tan\Phi)^2}{(1+P\tan\Phi)^2}\frac{1}{\sinh^2Z}\nonumber\\
=&1-\frac{P^2-(P\cos\Phi-\sin\Phi)^2}{(\cos\Phi+P\sin\Phi)^2}\nonumber\\
=&\frac{1}{(\cos\Phi+P\sin\Phi)^2}.
\end{align}
Equation \eqref{Npbh} then becomes
\begin{equation}
N'(\phi')=\frac{\cot\Phi P-1}{\sinh Z}.
\end{equation}
We now want to derive expression \eqref{radeqbh}. Since $\coth Z=\frac{P}{P\cos\Phi-\sin\Phi}$, we have $\sinh Z\cosh Z=\frac{P(P\cos\Phi-\sin\Phi)}{P^2-(P\cos\Phi-\sin\Phi)^2}$ and $\sinh^2 Z=\frac{(P\cos\Phi-\sin\Phi)^2}{P^2-(P\cos\Phi-\sin\Phi)^2}$, and we obtain that
\begin{align}
\frac{1}{\cosh X \cosh Z+\sinh Z}&=\sinh Z \frac{P^2-(P\cos\Phi-\sin\Phi)^2}{P(P\cos\Phi-\sin\Phi)(\cos\Phi+P\sin\Phi)+(P\cos\Phi-\sin\Phi)^2}\nonumber\\
&=\frac{\sinh Z}{P\cot\Phi-1},
\end{align}
which proves \eqref{radeqbh}, and where we have used that $\cosh X=P\cos\Phi+\sin\Phi$ since it can be seen that $P\cos\Phi+\sin\Phi>0$.

\section{Derivation of equation \eqref{awesome}}\label{app4}
Let us for concreteness assume that a black hole has formed. The formation of a conical singularity works similarly. We start by proving the following identity
\begin{equation}
\frac{G'}{G}=P.\label{GPGP}
\end{equation}
From \eqref{Gdefbh} and \eqref{Zbh} it is possible to derive that
\begin{equation}
G=\frac{\sqrt{M}|\sin\Phi|}{\sqrt{P^2\sin^2\Phi+2P\cos\Phi\sin\Phi-\sin^2\Phi}}.\label{Geq}
\end{equation}
We now have
\begin{equation}
\frac{G'}{G}=\frac{P\cos\Phi(\sin\Phi)'-PP'\sin^2\Phi-P'\cos\Phi\sin\Phi-P(\cos\Phi)'\sin\Phi}{P^2\sin^2\Phi+2P\cos\Phi\sin\Phi-\sin^2\Phi}.
\end{equation}
After using $(\sin\Phi)'\cos\Phi-(\cos\Phi)'\sin\Phi=\cos^2\Phi(\tan\Phi)'$, we can use equations \ref{ode1} and \ref{ode2} to eliminate the derivatives, and this then reduces precisely to \eqref{GPGP}.\\

Now it is easy to prove \eqref{awesome}. Using \eqref{Geq}, \eqref{GPGP}, $(G'/G)'=G''/G-(G')^2/G^2$ and the differential equation \eqref{ode1} we obtain from \eqref{Keq} that
\begin{align}
K^+_{\phi\phi}=&\frac{r}{2a_-}\left(\frac{-M}{G^2}-1\right)-\left(\frac{rG'(\phi)}{a_-G(\phi)}\right)'-\frac{r(G')^2}{2a_-G^2}\nonumber\\
=&\frac{r}{2a_-}\left((-P^2+1-2P\cot\Phi-1)-2P'-P^2\right)\nonumber\\
=&\frac{r}{2a_-}\left(-P^2-2P\cot\Phi-4\rho+2P^2+2P\cot\Phi-P^2\right)\nonumber\\
=&-\frac{r}{a_-}2\rho.
\end{align}
These manipulations are strictly not valid when $\rho$ is constant, but this special case can be done easily and we obtain the same result. The conical singularity case works analogously.

\end{document}